\documentclass{pasa}%

\usepackage{graphicx}
\usepackage{mathtools}
\usepackage{longtable}
\DeclarePairedDelimiter\ceil{\lceil}{\rceil}

\newcommand{\wrappedcell}[2][11.5em]{
\begin{minipage}[t]{#1}
\raggedright
#2
\end{minipage}}

\title[Hydra]{Hydra I: An extensible multi-source-finder comparison and cataloguing tool.}

\author[Boyce et al.]{M. M. Boyce,$^{1}$ A. M. Hopkins,$^{2}$  S. Riggi,$^{3}$  L. Rudnick,$^{4}$ M. Ramsay,$^{1}$ C. L. Hale,$^{5}$  J. Marvil,$^{6}$ M. Whiting,$^{7}$ P. Venkataraman,$^{8}$  C. P. O'Dea,$^{1}$ S. A. Baum,$^{1}$ Y. A. Gordon,$^{1, 9}$ A. N. Vantyghem,$^{1}$ M. Dionyssiou,$^{8}$ H. Andernach,$^{10}$ J. D. Collier,$^{11, 12}$ J. English,$^{1}$ B. S. Koribalski,$^{7, 12}$ D. Leahy,$^{13}$ M. J. Michałowski,$^{14}$ S. Safi-Harb,$^{1}$ M. Vaccari,$^{15, 16}$ 
E. Alexander,$^{17}$ M. Cowley,$^{18,19}$ A. D. Kapinska,$^{6}$ A. S. G. Robotham,$^{20}$ H. Tang.$^{21}$
\affil{$^{1}$Department of Physics and Astronomy, University of Manitoba, 30A Sifton Road, Winnipeg, MB R3T 2N2, Canada}%
\affil{$^{2}$Australian Astronomical Optics, Macquarie University,
105 Delhi Rd, North Ryde, NSW 2113, Australia}
\affil{$^{3}$INAF, Osservatorio Astrofisico di Catania
Via S. Sofia 78, 95123, Catania, Italy}
\affil{$^{4}$Minnesota Institute for Astrophysics, School of Physics and Astronomy, University of Minnesota, 116 Church Street SE, Minneapolis, MN 55455, USA}
\affil{$^{5}$School of Physics and Astronomy, University of Edinburgh, Institute for Astronomy, Royal Observatory, Blackford Hill, Edinburgh EH9 3HJ, UK}
\affil{$^6$National Radio Astronomy Observatory, P.O. Box O, Socorro, NM 87801, USA}
\affil{$^{7}$Australia Telescope National Facility, CSIRO Astronomy and Space Science, PO Box 76, Epping, NSW 1710, Australia}
\affil{$^{8}$Dunlap Institute for Astronomy and Astrophysics, University of Toronto, 50 St. George Street, Toronto, ON M5S 3H4, Canada}
\affil{$^{9}$Department of Physics, University of Wisconsin-Madison, Madison, WI 57306, USA}
\affil{$^{10}$Departamento de Astronom{\'{i}}a, DCNE, Universidad de Guanajuato, Callej\'on de Jalisco s/n, Guanjuato, CP 36023, GTO, Mexico}%
\affil{$^{11}$Inter-University Institute for Data Intensive Astronomy (IDIA), Department of Astronomy, University of Cape Town, Private Bag X3, Rondebosch, 7701, South Africa}
\affil{$^{12}$School of Science, Western Sydney University, Locked Bag 1797, Penrith, NSW 2751, Australia}
\affil{$^{13}$Department of Physics and Astronomy, University of Calgary, 2500 University Dr. NW, Calgary, AB T2N 1N4, Canada}
\affil{$^{14}$Astronomical Observatory Institute, Faculty of Physics, Adam Mickiewicz University, ul.~S{\l}oneczna 36, 60-286 Pozna{\'n}, Poland}
\affil{$^{15}$Inter-University Institute for Data Intensive Astronomy (IDIA), Department of Physics and Astronomy, University of the Western Cape, Robert Sobukwe Road, 7535 Bellville, Cape Town, South Africa}
\affil{$^{16}$INAF - Istituto di Radioastronomia, via Gobetti 101, 40129 Bologna, Italy}
\affil{$^{17}$Jodrell Bank Centre for Astrophysics, Department of Physics and Astronomy, University of Manchester, Manchester, UK}
\affil{$^{18}$School of Chemistry and Physics, Queensland University of Technology, Brisbane, QLD 4000, Australia}
\affil{$^{19}$Centre for Astrophysics, University of Southern Queensland, West Street, Toowoomba, QLD 4350, Australia}
\affil{$^{20}$ICRAR, M468, University of Western Australia, Crawley, WA 6009, Australia}
\affil{$^{21}$Department of Astronomy, Tsinghua University, Beijing 100084, China}
}

\jid{PASA}
\doi{10.1017/pas.\the\year.xxx}
\jyear{\the\year}

\usepackage{aas_macros}
\usepackage{hyperref} 
\hypersetup{colorlinks,citecolor=blue,linkcolor=blue,urlcolor=blue}

\usepackage{soul}

\begin{document}

\begin{frontmatter}
\maketitle

\begin{abstract}
The latest generation of radio surveys are now producing sky survey images containing many millions of radio sources. In this context it is highly desirable to understand the performance of radio image source finder (SF) software and to identify an approach that optimises source detection capabilities. We have created Hydra to be an extensible multi-SF and cataloguing tool that can be used to compare and evaluate different SFs. Hydra, which currently includes the SFs Aegean, Caesar, ProFound, PyBDSF, and Selavy, provides for the addition of new SFs through containerisation and configuration files. The SF input RMS noise and island parameters are optimised to a 90\% ``percentage real detections'' threshold (calculated from the difference between detections in the real and inverted images), to enable comparison between SFs. Hydra provides completeness and reliability diagnostics through observed-deep ($\mathcal{D}$) and generated-shallow ($\mathcal{S}$) images, as well as other statistics. In addition, it has a visual inspection tool for comparing residual images through various selection filters, such as S/N bins in completeness or reliability. The tool allows the user to easily compare and evaluate different SFs in order to choose their desired SF, or a combination thereof. This paper is part one of a two part series. In this paper we introduce the Hydra software suite and validate its $\mathcal{D/S}$ metrics using simulated data. The companion paper demonstrates the utility of Hydra by comparing the performance of SFs using both simulated and real images.
\end{abstract}

\begin{keywords}
methods: data analysis -- radio continuum: general -- techniques: image processing
\end{keywords}
\end{frontmatter}


\section{Introduction}
\label{sec:intro}
With the advent of new facilities, radio surveys are becoming larger and deeper, providing fields rich in sources, in the tens of millions \citep{norris_2017}, and delivering data at increasing rates, in the hundreds of gigabytes per second \citep{whiting_2012}. The Evolutionary Map of the Universe \citep[EMU,][]{norris_2011,norris_2021} is expected to detect up to 40~million sources, expanding our knowledge in areas such as galaxy evolution and star formation. This outstrips surveys like the Karl G. Jansky Very Large Array (JVLA, or VLA) Sky Survey \citep[VLASS,][]{lacy_2020,gordon_2021} and the Rapid Australian Square Kilometer Array (SKA) Pathfinder (ASKAP) Continuum Survey \citep[RACS,][]{mcconnell_2020,hale_2021} by a factor of up to 30. Furthermore, the Variable and Slow Transients \citep[VAST,][]{banyer_2012,murphy_2013,murphy_2021} survey, operating at a cadence of $5s$, surpasses VLASS transient studies by several orders of magnitude, opening up areas of variable and transient research: \textit{e.g.}, flare stars, intermittent pulsars, X-ray binaries, magnetars, extreme scattering events, interstellar scintillation, radio supernovae, and orphan afterglows of gamma-ray bursts \citep{murphy_2013,murphy_2021}. This places complex requirements on source finder (SF) software in order to reliably handle compact,\footnote{Herein, compact refers to point sources.} extended, complex, and faint or diffuse sources, along with demands for high data throughput for radio transients  \citep[\textit{e.g.},][]{hancock_2012,hopkins_2015,riggi_2016,hale_2019,boyce_2020,bonaldi_2021}. No current SF fits all of these requirements. 

Hydra is an attempt to get the best of all worlds: it is an extensible multi-SF comparison and cataloguing tool, which allows users to choose the appropriate SF for a given survey, or take advantage of its collectively rich statistics by combining results. The Hydra software suite\footnote{Hydra is available, along with the data products presented in this paper, by navigating through the CIRADA portal at \url{https://cirada.ca}. A more permanent home is expected, once VLASS data product development has been completed.} currently includes Aegean \citep{hancock_2012,hancock_2018}, Caesar \citep[Compact And Extend Source Automated Recognition,][]{riggi_2016,riggi_2019}, ProFound \citep{robotham_2018,hale_2019}, PyBDSF \citep[Python Blob Detector and Source Finder,][]{mohan_2015}, and Selavy \citep{whiting_2012}.

This paper is part one of a two part series, (referred to hereafter as Papers~I and II). Here we provide a brief overview of SFs, relevant to our implementation of Hydra. This is then followed by a description of the Hydra software suite. The software produces new metrics for handling real source components (or sources, herein), such as, completeness ($\mathcal{C}$) and reliability ($\mathcal{R}$), based on sources detected in a shallow ($\mathcal{S}$) image (\textit{e.g.}, a real image with artificial noise added) wherein real (sometimes referred to as ``deep'' or $\mathcal{D}$) image detections are considered as true sources. We use simulated data, where the true sources are known, to validate these metrics. In Paper~II we use the simulated images along with real data to evaluate the performance of the six different SFs included with Hydra. A preliminary discussion on SF performance is presented in this paper.

\section{Source Finders}
\label{sc:source_finders}
The growing sizes and data rates of modern radio surveys have increased the need for automated source finding tools with fast processing speeds, and high completeness and reliability. One impetus for this came through the ASKAP EMU source finding data challenge \citep{hopkins_2015}, which explored a community-submitted set of eleven SFs: Aegean \citep{hancock_2012}, Astronomical Point source EXtractor \citep[APEX,][]{makovoz_2005}, \textsc{blobcat} \citep{hales_2012}, Curvature Threshold Extractor \citep[CuTEx,][]{molinari_2011}, Duchamp \citep{whiting_2012b}, IFCA (International Federation of Automation Control) Biparametric Adaptive Filter \citep[BAF,][]{lopez_2012} / Matched Filter \citep[MF,][]{lopez_2006}, PyBDSF \citep{mohan_2015}, Python Source Extractor \citep[PySE,][]{spreeuw_2010,swinbank_2015}, Search and Destroy (SAD, \cite{condon_1998}, with an honourable mention of its variant HAPPY, \cite{white_97}), Selavy \citep[with Duchamp at its core,][]{whiting_2012}, Source Extractor \citep[SExtractor,][]{bertin_1996}, and SOURCE\_FIND \citep[Arcminute Microkelvin Imager (AMI) pipeline,][]{amic_2011}. More recent SFs include Caesar \citep{riggi_2016,riggi_2019} and ProFound \citep{robotham_2018,hale_2019,boyce_2020}. APEX, CuTEx, ProFound, and SExtractor have their origins in optical astronomy. Our focus will be on 2D SFs, such as those above, although there are also 3D packages like SoFiA \citep[Source Finding Application,][]{serra_2015,koribalski_2020,westmeier_2021} optimised for detecting line emission in data cubes, which can also function as 2D SFs.

By and large there is no ``one SF fits all'' solution. Each is typically optimised for specific tasks \citep{hopkins_2015,hale_2019,bonaldi_2021}. In the broadest sense, there are SFs designed to handle sources that are compact, or extended and diffuse, see Table~\ref{tb:isf_characteristics}. They also have their specializations: \textit{e.g.}, \textsc{blobcat} for linear polarisation data \citep{hales_2012}, Duchamp for HI observations \citep{whiting_2012b}, CuTEx for images with intense background fluctuations \citep{molinari_2011}, and PySE for transients \citep{fender_2007,haarlem_2013}. There are also \textit{``Next Generation''} (\textsc{NxGen}) SFs (see Table \ref{tb:isf_characteristics}), which utilize multiple processors for handling high data throughput \citep{hancock_2012,riggi_2016,whiting_2012}. Qualitatively different types of source-finding and characterisation tools are being developed that use machine learning approaches \cite[\textit{e.g.,}][]{bonaldi_2021,lao21,magro_2022}, as well as citizen science approaches to classifying radio sources \cite[\textit{e.g.,}][]{banfield15,alger18}, although it is beyond the scope of Hydra to attempt to incorporate all such efforts.

\begin{table}[htb]
   \caption{SF general design characteristics \citep[\textit{re.},][]{hopkins_2015,hale_2019,bonaldi_2021}. \textsc{NxGen} indicates multiprocessing capabilities.}
   \centering
   {\small
   \begin{tabular}{@{\;}l@{\;}c@{\;}c@{\;}@{\;}c@{\;}@{\;}c@{\;}@{\;}c@{\;}}
        \hline\hline
        SF    & \multicolumn{3}{c}{Source Type}      & \textsc{NxGen}\\
                         &   Compact  &  Extended  & Diffuse    \\
        \hline
        Aegean           & \checkmark &            &            & \checkmark \\
        APEX$^a$         & \checkmark &            &            &            \\
        \textsc{blobcat} & \checkmark & \checkmark &            &            \\
        Caesar           & \checkmark & \checkmark & \checkmark & \checkmark \\
        CuTEx$^a$        & \checkmark &            &            &            \\
        IFCA BAF/MF      &            & \checkmark &            &            \\
        Selavy           & \checkmark &            &            & \checkmark \\
        ProFound$^a$     & \checkmark    & \checkmark & \checkmark &            \\
        PyBDSF           & \checkmark &   \checkmark     &            &            \\
        PySE             & \checkmark &            &            &            \\
        SAD              & \checkmark &            &            &            \\
        SExtractor$^a$   &            & \checkmark &            &            \\
        SOURCE\_FIND     &            & \checkmark &            &            \\
        \hline\hline
        \multicolumn{5}{l}{$^a$Optical SF.}
   \end{tabular}
   }
   \label{tb:isf_characteristics}
\end{table}

In general, SFs typically analyse an image in 3 stages: (1)~background and noise estimation, (2)~island detection, and (3)~component modelling. 

For the background estimation most SFs used in radio astronomy such as Aegean, PyBDSF, and Selavy, tend to use a sliding box method, where background noise estimates are calculated at a specific location using neighbouring pixels within a given box size, and estimated again for adjacent locations based on the sliding-step size. It is important that the box size be set so as not to be too small around bright sources, which would overestimate the background noise, or too large, so as to wash out any varying background structure that is important for reliable detection of faint sources \citep[\textit{e.g.},][]{huynh_2012}. This is discussed further below in \S\,\ref{sc:typhon}.

Background noise estimation can be performed through various metrics such as the inter-quartile range (IQR) used by Aegean \citep[with median background and IQR noise spread,][]{hancock_2012}, mean background ($\mu$) and RMS noise ($\sigma$) used by PyBDSF \citep{mohan_2015}, or median background and Mean Absolute Deviation From the Median \citep[MAD, or MADFM herein: \textit{e.g.},][]{riggi_2016,hopkins_2015} noise used by Selavy \citep[which also has a $\mu$/$\sigma$ option,][]{whiting_2012}. SExtractor, on the other hand, uses $\kappa.\sigma$-clipping and mode estimation \citep[see \S~\ref{sc:homados};][]{costa_1992,bertin_1996,huynh_2012,akhlaghi_2015,riggi_2016} over the entire image; while PySE performs $\sigma$-clipping locally \citep[see][]{hopkins_2015}. ProFound, an optical SF shown to be useful for radio images \citep{hale_2019}, also uses a $\sigma$-clipping schema \citep[\textit{via.} the \textsc{MakeSkyGrid} routine,][]{robotham_2018}. Caesar provides several options: $\mu/\sigma$, median/MADFM, biweight and $\sigma$-clipped estimators \citep{riggi_2016}. The final stage typically involves bicubic interpolation to obtain the background noise estimates as a function of pixel location. It is important that these estimates are optimal as they have a significant effect on SF performance \citep{huynh_2012}.

There are various methods for island detection within an image. Perhaps the simplest is thresholding, in which the pixel with the highest flux is chosen along with neighboring pixels down to some threshold above the background noise, defining an island.  Variants of this method are used by Duchamp \citep{whiting_2012}, ProFound \citep{robotham_2018}, Selavy \citep{whiting_2012b}, and SExtractor \citep{bertin_1996}. Once the initial set of islands are chosen, they are sometimes then grown down to a lower threshold according to certain rules. For instance, ProFound uses a Kron/Petrosian-like dilation kernel, \textit{i.e.}, it uses an island-shaped aperture \citep{kron_1980,petrosian_1976} to grow the islands according to a surface brightness profile, in an iterative process, until the desired profile or lower threshold limit is reached \citep{robotham_2018}. It then separates out the islands into segments, through a watershed deblending technique.\footnote{The term ``watershed'' refers to drainage basins formed from streams running between mountains (islands), during a rainfall, following the steepest descent \citep{beucher_1979}.} Another method is flood-fill, wherein islands are seeded above some threshold and then grown down to a lower threshold, according to a set of rules. Aegean \citep{hancock_2012}, \textsc{blobcat} \citep{hales_2012}, Caesar \citep{riggi_2016}, PyBDSF \citep{mohan_2015}, and PySE \citep{swinbank_2015} use variations on this theme. 

The component extraction phase is perhaps the most varied in terms of modelling. The simplest is the top down raster-scan within an island to find flux peaks given some step size, or tolerance level. This method is utilized by Duchamp \citep{whiting_2012}, and, in turn, is also employed by Selavy. These peaks are then fitted by elliptical Gaussians producing a component catalogue. The choice of elliptical Gaussians is motivated by the fact that point sources, or sources that are only very slightly extended, are well-modelled in this way as it corresponds to the shape of the telescope's synthesised beam. More complex source structures, on the other hand, tend to be poorly fit by this choice, leading to variations in fitting approach. Some SFs, for example, use multiple Gaussians to fit to an island, using various criteria. PyBDSF \citep{mohan_2015} and PySE \citep{spreeuw_2010,swinbank_2015} fall into this category.

There are also a class of SFs that use curvature maps to determine radio source components. Aegean searches for local maxima within an island which in turn are fitted by Gaussians, constrained by negative curvature valleys \citep{hancock_2012}. Caesar is rather unique in that it first searches for peaks and then uses watershed deblending to create sub-islands, for seeding and constraining Gaussian fits, respectively \citep{riggi_2016}. Extended sources are then extracted from the resulting residual image, using wavelet-transform, saliency, hierarchical-clustering, or active-contour filtering. Consequently, Caesar is capable of extracting extended sources with complex structure.

The aforementioned SF algorithms are just the tip of the iceberg of possibilities \citep[\textit{c.f.},][]{hancock_2012,hopkins_2015,wu_2018,lukic_2019,sadr_2019,bonaldi_2021,magro_2022}. In our initial implementation of Hydra we have chosen to explore a representative set of commonly used SFs: Aegean, Caesar, ProFound, PyBDSF, and Selavy.

\section{Hydra}
\label{sc:hydra}
Hydra is a software tool capable of running multiple SFs. It is extensible, in that other SFs can be added in a containerised fashion by following a set of straightforward template-like rules. It provides diagnostic information such as $\mathcal{C}$ and $\mathcal{R}$. Statistical analysis can be based on injected ($\mathcal{J}$) source catalogues from simulated images or on real $\mathcal{D}$-images used as ground-truths for detections in their $\mathcal{S}$ counterparts. Hydra is innovative in that it minimises the False Detection Rate \citep[FDR,][]{whiting_2012b} of the SFs by adjusting their detection threshold and island growth (and optionally RMS box) parameters. This is an essential step in automation, especially when dealing with large surveys such as EMU \citep{norris_2017,norris_2021}.

\subsection{The Hydra Software Suite}
\begin{quote}
\textit{``Upon Heracles shield wrought Homados (Tumult), the din of battle noise, and riding alongside Cerberus, the unruly master of mayhem. Only the wrath of Cerberus's father Typhon, a controlling force, can temper their chaos. And hitherto, Typhon's son Hydra, was tasked with bringing the chaos to bear fruit, while his mother Echidna, a hidden force, plucked the fruit from the vines to make wine.'' (Inspired from \citeauthor{powell_2017}, \citeyear{powell_2017}, and \citeauthor{buxton_2016}, \citeyear{buxton_2016}.)}
\end{quote}

\begin{figure}[htb!]
\begin{center}
\includegraphics[width=\columnwidth]{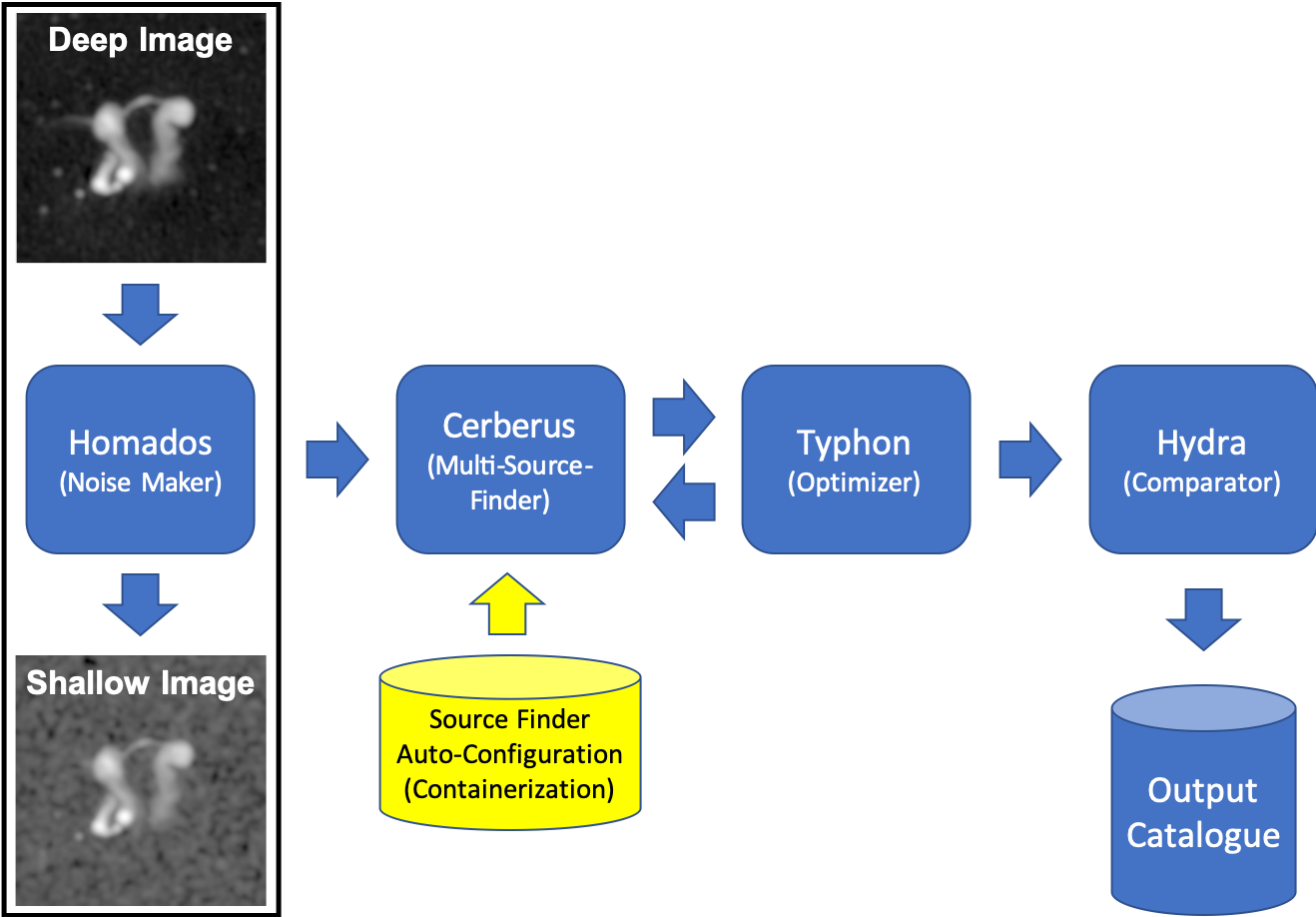}
\caption{High level schematic representation of the Hydra software suite workflow. Homados provides $\mathcal{D}$ and $\mathcal{S}$-image channels for simulated/real images \citep[``dancing ghosts'' example image, see][]{norris_2021}. Each channel is run separately through the Typhon optimiser, which uses the SF interface provided by Cerberus. Hydra coordinates all of these activities, building catalogues and compiling statistics at the end of the process. \label{fg:hydra_suite}} 
\end{center}
\end{figure}

\noindent Figure~\ref{fg:hydra_suite} shows an overview of the Hydra software suite, which consists of the following software components: Homados, Cerberus, Typhon, and Hydra. Homados is used for providing image statistics such as $\mu/\sigma$, and image manipulation such as inversion and adding noise. Cerberus is an extensible multi-SF software suite. It currently includes Aegean \citep{hancock_2012,hancock_2018}, Caesar \citep{riggi_2016,riggi_2019}, ProFound \citep{robotham_2018,hale_2019}, PyBDSF \citep{mohan_2015}, and Selavy \citep{whiting_2012}. Typhon is a tool for optimising the SF parameters and then producing output catalogues. It uses Homados and Cerberus to do this task. Hydra is the main tool which uses Typhon to produce data products, including catalogues, residual images, and region files.\footnote{Hydra refers to both the software suite and the software tool \texttt{hydra.py}. There should be no source of confusion in this regard, as only \texttt{hydra.py} is used to create the data products.} Echidna is a planned catalogue stacking and integration tool, to be added to Hydra.

\subsubsection{Homados}
\label{sc:homados}
The main purpose of Homados is to add noise to images. We shall often refer to the original image, as the $\mathcal{D}$-image, and the noise-added image, as the $\mathcal{S}$-image. These ``deep-shallow'' ($\mathcal{DS}$) image pairs can be used to create statistics such as $\mathcal{DS}$-completeness ($\mathcal{C_{DS}}$) and $\mathcal{DS}$-reliability ($\mathcal{R_{DS}}$), based on the assumption that the sources detected in the $\mathcal{D}$-image are real. These statistics are used for real images, where the source inputs are unknown. 

An $\mathcal{S}$-image is created by adding to the $\mathcal{D}$-image a Gaussian noise map that has been convolved with the corresponding synthesised beam (\textit{i.e.}, \texttt{BMIN}, \texttt{BMAJ}, and \texttt{BPA}). The noise map is created with mean noise, $\mu_{image}$, and RMS noise, $n\sigma_{image}$ ($\equiv\sigma_{noise\;\;map}$), where $n$ is the desired noise level (\textit{i.e.}, factor), and $\mu_{image}$ and $\sigma_{image}$ are obtained from the $\mathcal{D}$-image using $\sigma$-clipping \citep{akhlaghi_2015}. This is then convolved with the synthesised beam, from which its RMS noise, $\sigma_{convolved}$, is computed. For convergence, this process is repeated using the convolved image as input, but with $n$ replaced by $\sigma_{noise\;\;map}\,n/\sigma_{convolved}$. The final convolved image is then added to the $\mathcal{D}$-image, obtaining the $\mathcal{S}$-image.

\begin{figure}[htb!]
\begin{center}
\includegraphics[width=\columnwidth]{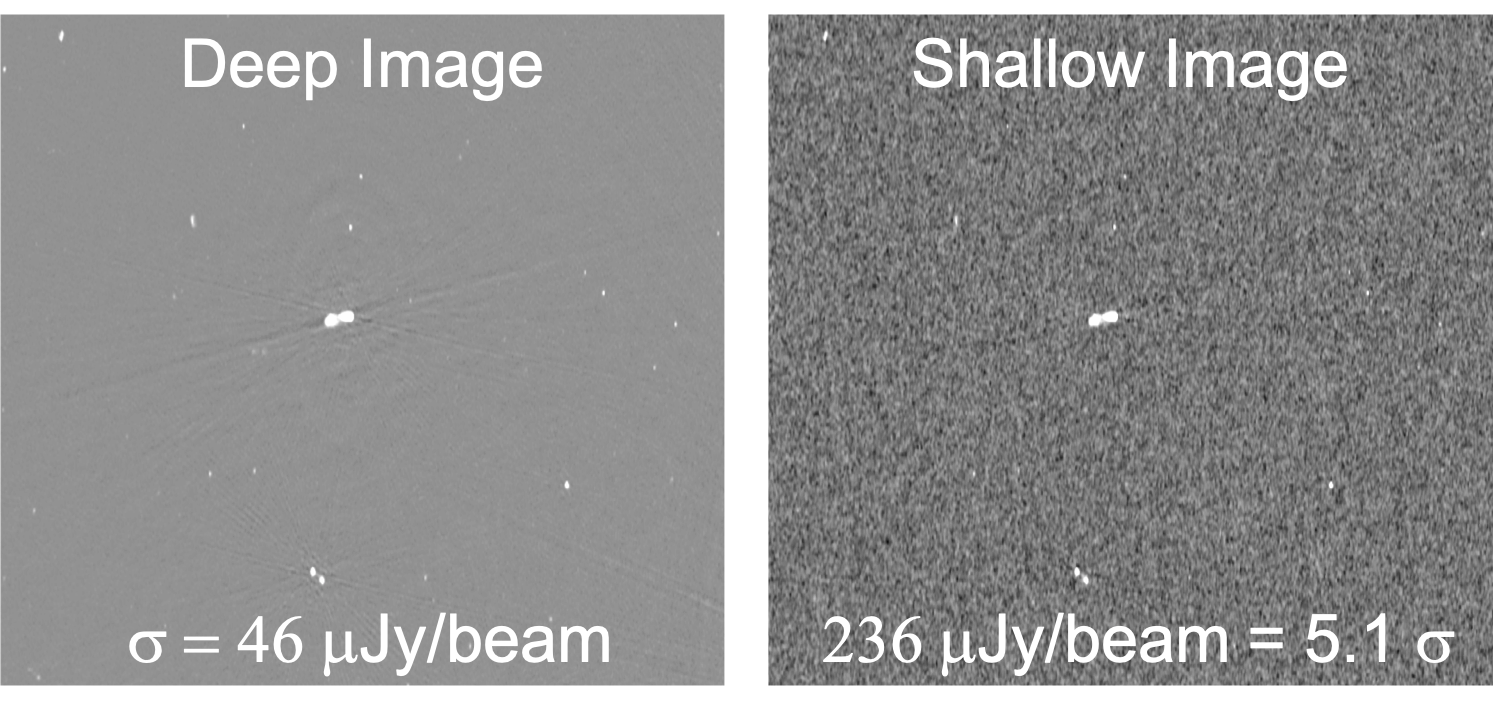}
\caption{Homados $\mathcal{S}$-image generation example, using an image cutout sample from an ATLAS CDFS DR1 $2.2^{\circ}\times2.7^{\circ}$ tile   \citep{norris_2006}. The figures show $\mathcal{D}$ (left) and $\mathcal{S}$ (right) images, zoomed in. The noise level scale factor, $n$, was set to 5 to generate the shallow image.\label{fg:homados_atlas_example}}
\end{center}
\end{figure}

Figure~\ref{fg:homados_atlas_example} shows an example of an $\mathcal{S}$-image generated by Homados from an Australia Telescope Large Area Survey (ATLAS) Chandra Deep Field South (CDFS) Data Release 1 (DR1) tile \citep{norris_2006}. The noise level was scaled by a factor of $n=5$. This factor is assumed for the $\mathcal{S}$-image generation in the rest of this paper. 

In addition, Homados uses $\sigma$-clipping to compute image statistics, such as, $m$ (median), $\mu$ (mean), $\sigma$ (RMS), $I_{min}$ (minimum pixel value), and $I_{max}$ (maximum pixel value). It also does image inversion for FDR calculations.

\subsubsection{Cerberus}
\label{sc:cerberus}
\begin{figure}[htb!]
\begin{center}
\includegraphics[width=\columnwidth]{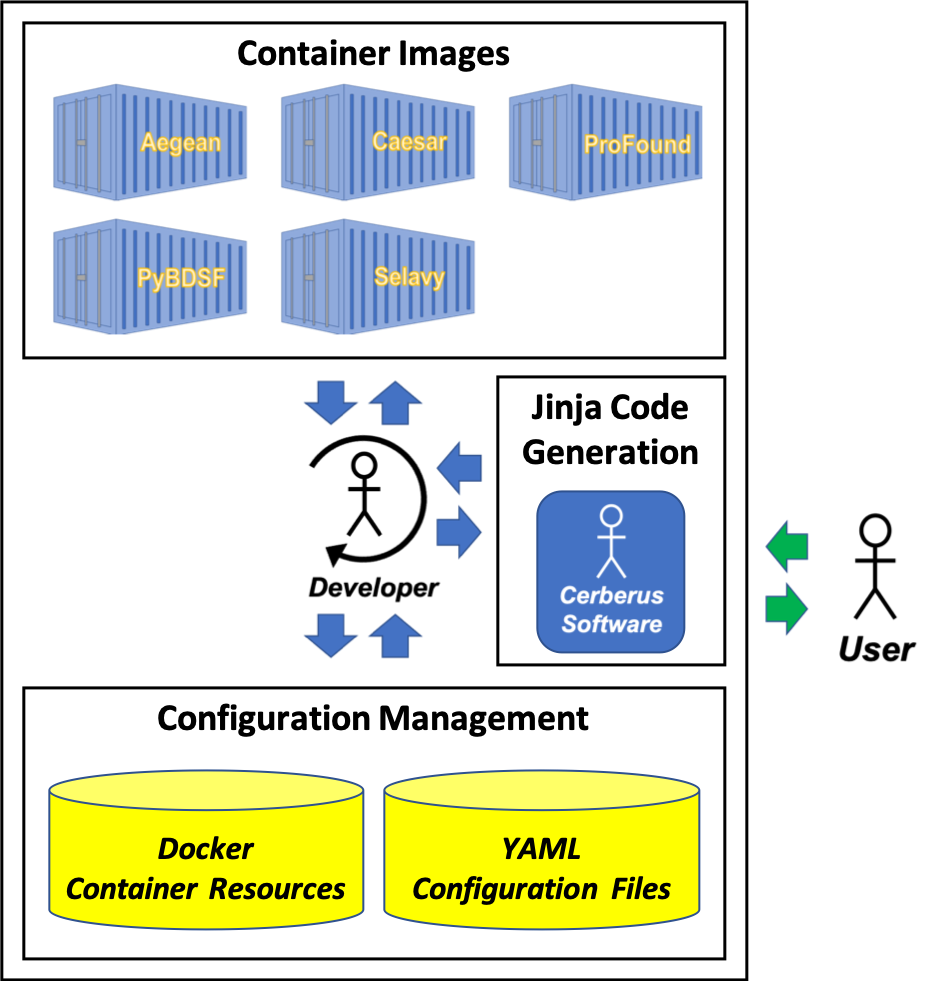}
\caption{Cerberus code generation workflow.} \label{fg:Cerberus_Code_Generation}
\end{center}
\end{figure}

Cerberus is an extensible interface for running SF modules within the Hydra software suite. It currently supports Aegean, Caesar, ProFound, PyBDSF, and Selavy, as indicated by its command-line interface.\footnote{The command-line interface for all Hydra tools is standardised using Click, \url{https://click.palletsprojects.com}. Click allows direct interface calls within a script, through its \texttt{standalone\_mode} (\texttt{=True}) flag, which allows for interoperability between Hydra tools while preserving their user interfaces.}
{\footnotesize
\begin{verbatim}
$ python cerberus.py --help
Usage: cerberus.py [OPTIONS] COMMAND [ARGS]...

  Runs a collection of source finder modules.

Options:
  --help  Show this message and exit.

Commands:
  process   Use multiple source finders.
  aegean    Use aegean source finder.
  caesar    Use caesar source finder.
  profound  Use profound source finder.
  pybdsf    Use pybdsf source finder.
  selavy    Use selavy source finder.
$
\end{verbatim}}
New modules are added through code generation, using Jinja template-code\footnote{Jinja (\url{https://jinja.palletsprojects.com}) is similar to Django (\url{https://www.djangoproject.com}) templates for dynamically creating webpages, but can also be applied to software.} in conjunction with Docker\footnote{Docker (\url{https://www.docker.com}) is used to hide the complexity of installing and operating SFs, by wrapping them inside their own mini-operating system environment.} and YAML\footnote{YAML (\url{https://yaml.org}) configuration files are used to store Python-like data structures, but in human readable form.} configuration files. The workflow is as follows:
\begin{itemize}
    \item Create a containerised SF wrapper:
    \begin{itemize}
        \item Create a SF wrapper script
        \item Create a Docker build file wrapper
        \item Update the master docker-compose build file
        \item Build the container image
    \end{itemize}
    \item Update the \texttt{cerberus.py} script:
    \begin{itemize}
        \item Create a YAML metadata file
        \item Update the master YAML metadata file
        \item Run the Jinja script generator tool
    \end{itemize}
    \item Test the Hydra software suite
    \item Update the Git\footnote{\url{https://git-scm.com}} repository
\end{itemize}
\noindent Figure~\ref{fg:Cerberus_Code_Generation} summarises this high-level workflow: containers for each SF are shown under Container Images and the Docker and YAML configuration files are shown under Configuration Management. The developer must follow a fixed set of rules when adding a new SF, in order for the Jinja template-driven script generator to update Cerberus. (For the purpose of reproducibility, Appendix~\ref{ap:cerberus_detail} provides architectural design notes, using Aegean as an example.) All of this is transparent to the user, who has access to a simple interface, so one does not have to be an expert at using SFs in order to use Hydra.

Hydra's modular design requires that the user has access to the key elements of a SF's interface; in particular, access to its ``RMS-like'' and ``Island-like'' parameters. In the case of Aegean, for example, this would be \texttt{seedclip} and \texttt{floodclip} \citep{hancock_2012}, respectively. It is important to note that the parameters are not necessarily equivalent between SFs;\footnote{Appendix~\ref{ap:source_finder_notes} delineates these differences.\label{fn:delineates}} regardless, they do affect thresholding and island formation. Consequently, they have the strongest influence on FDR calculations. Table~\ref{tb:rms_isl_pars} summarises the parameters for the currently supported SFs. These parameters are used by Typhon to baseline the SFs, by minimizing their FDRs.

\begin{table*}[htb!]
\caption{Cerberus RMS and Island parameter definitions in units of $\sigma$ with respect to the background, with soft constraint $\sigma_{Island}<\sigma_{RMS}$.} 
\centering
\begin{tabular*}{\textwidth}{@{$\;\;$}l\x l\x c\x c\x l\x c\x c@{$\;\;$}}
\hline \hline
 Source & \multicolumn{3}{c}{RMS Parameter} &  \multicolumn{3}{c}{Island Parameter} \\
 Finder & Name & Default  & Description &  Name  &  Default & Description \\
\hline
&&&&&&\\ 
 Aegean$^a$ & \texttt{seedclip} & 5.0$\;\;$ & \wrappedcell{The clipping value for seeding islands.} & \texttt{floodclip} & 4.0 & \wrappedcell{The clipping value for growing islands.}\\ &&&&&&\\ 
 Caesar$^b$ & \texttt{seedThr} & 5.0$\;\;$ & \wrappedcell{Blob finding threshold.} & \texttt{mergeThr} & 2.6 & \wrappedcell{Blob growth threshold.}\\
&&&&&&\\ 
 ProFound$^c$ & \texttt{skycut} & 2.8 & \wrappedcell{Island threshold.} 
          & \texttt{tolerance} & 4.0 & \wrappedcell{Defines island separation height.}\\
&&&&&&\\ 
 PyBDSF$^d$ & \texttt{thresh\_pix} & 5.0$\;\;$ & \wrappedcell{Source detection threshold.} 
        & \texttt{thresh\_isl} & 3.0 & \wrappedcell{Threshold for the island boundary.}\\
&&&&&&\\ 
 Selavy$^e$ & \texttt{snrCut} & 4.0$\;\;$ & \wrappedcell{Detection threshold.} 
       & \texttt{growthCut} & 3.0 & \wrappedcell{Threshold value to grow detections down to.}\\
&&&&&&\\ 
\hline \hline
\end{tabular*}\label{tb:rms_isl_pars}
\medskip
\tabnote{$^a$\url{https://github.com/PaulHancock/Aegean/wiki/Simple-usage}}
\tabnote{$^b$\url{https://caesar-doc.readthedocs.io/en/latest/usage/app\_options.html\#input-options}}
\tabnote{$^c$\url{https://cran.r-project.org/web/packages/ProFound/ProFound.pdf}}
\tabnote{$^d$\url{https://pybdsf.readthedocs.io/en/latest/process_image.html}}
\tabnote{$^e$\url{https://www.atnf.csiro.au/computing/software/askapsoft/sdp/docs/current/analysis/selavy.html}}
\end{table*}

Hydra also requires that SF modules provide optional RMS box and step size parameters, even if they are dummies. Some SF software manuals recommend these parameters be externally optimized, under certain conditions. PyBDSF is an example of such a case.\footnote{See \texttt{rms\_box} discussion at URL in footnote $d$ of Table~\ref{tb:rms_isl_pars}.} Regardless, this is also a good way of baselining \citep[\textit{i.e.} calibrating,][]{huynh_2012,riggi_2016} SFs for comparison purposes.

\subsubsection{Typhon}
\label{sc:typhon}
Typhon is a tool for optimising the SFs to a standard baseline that can be used for comparison purposes. We have adopted the Percent Real Detections (PRD) metric, as used by \cite{hale_2019} in a comparative study of Aegean, ProFound, and PyBDSF: \textit{i.e.},

\begin{equation}
    \mbox{PRD} = \frac{N_{image}-N_{inv.\;image}}{N_{image}}\,100
    \label{eq:prd}
\end{equation}

\noindent where $N_{image}$ is the number of detections in the original image, and $N_{inv.\;image}$ is the number of detections in the inverted image.\footnote{\textit{i.e.}, negative pixel values.} Basically, if one assumes the image noise is predominately Gaussian, then the peaks detected in the inverted image should statistically match the noise-peaks detected in the non-inverted image. Thus the FDR can be reduced by optimising the PRD. This approach is not suitable for non-Gaussian (or non-symmetric) noise properties, such as the Ricean noise distribution in polarisation images.

\begin{figure*}[htb!]
\begin{center}
\includegraphics[width=\columnwidth]{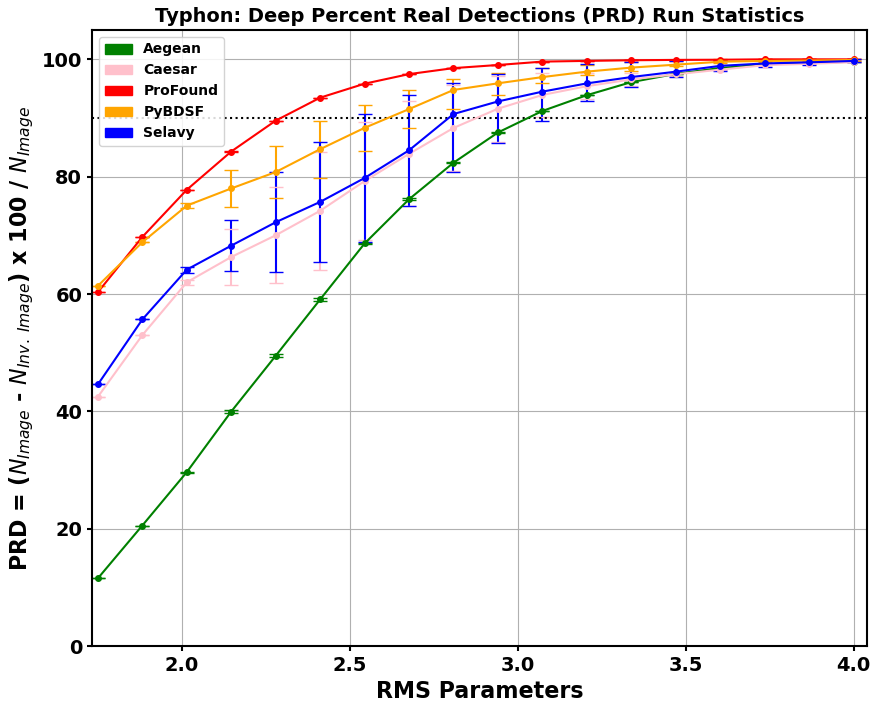}
\includegraphics[width=\columnwidth]{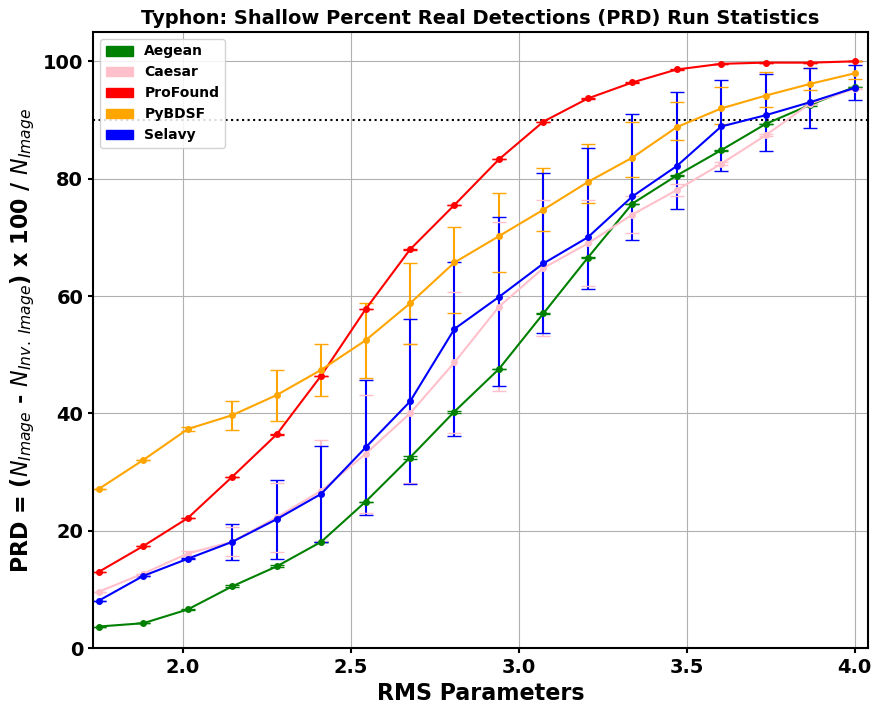}
\caption{Example Typhon PRD of a $2^\circ\times2^\circ$ simulated $\mathcal{D}$-image (left) and its corresponding $\mathcal{S}$-image (right). The variation in the PRD with the RMS parameters (in $\sigma$ units) is represented along the horizontal axis, and the variation in the PRD with the island parameters is represented by the error bars. The data points represent average values: \textit{i.e.}, Aegean and ProFound indicate the true shape of the curves, due their insensitivity to their island parameters. The SF parameters are listed in Table~\ref{tb:rms_isl_pars}. The dotted horizontal lines indicate the 90\% PRD levels. \label{fg:typhon_prd_example}}
\end{center}
\end{figure*}

Typhon uses the RMS and island parameters to optimise the PRD for each SF. Figure~\ref{fg:typhon_prd_example} shows Typhon generated PRD curves for a $2^\circ\times2^\circ$ simulated $\mathcal{D}$-image along with its corresponding $\mathcal{S}$-image. Typhon identifies the optimal parameters to be those that correspond to the 90\% PRD threshold. This threshold is motivated by the desire to use the knee of the PRD curve, whose position appears to be scale-invariant above a certain image size. Although the shape of the curve is not always guaranteed to be smooth, this crude method appears to be quite effective at framing the region of interest around the desired 90\% PRD. The 90\% to 98\% PRD range has been investigated and the former threshold seems to provide reasonable results. \cite{hale_2019} use a 98\% PRD to baseline their SFs, beyond which the detection rate degrades. At that cutoff, however, we tend to find a non-scale-invariant increase in the RMS threshold with image size.

Typhon uses the image statistics output from Homados to determine the RMS parameter range over which to optimise the PRD:
\begin{equation}
    1.5\sigma\le\mbox{RMS}\le\mbox{RMS}_{max}\,,\label{eq:rms_pars}
\end{equation}
where
\begin{equation}
    \mbox{RMS}_{max} = \ceil*{\frac{I_{max}-\mu}{\sigma}}\sigma\,.
    \nonumber
    \label{eq:rms_max}
\end{equation}
with $\mu$, $\sigma$, and $I_{max}$ determined through $\sigma$-clipping (\textit{re.} \S~\ref{sc:homados}). The $1.5\sigma$ lower limit is where the FDR starts to degrade. In general, Typhon samples the PRD from high values to low values in the RMS parameter, while varying the island parameter at each step, until the 90\% threshold is reached.

The island parameters are SF specific, and are typically defined over a finite range. Table~\ref{tb:island_ranges} shows the parameter ranges used by Hydra, which are stored in its Configuration Management (Figure~\ref{fg:Cerberus_Code_Generation}). Typhon uses this information along with the constraint $\sigma_{island}<\sigma_{RMS}$ (otherwise, $\sigma_{island}=0.999\,\sigma_{RMS}$), as it searches the parameter space.

\begin{table}[htb!]
\caption{Configured island parameters.}
\centering
\begin{tabular}{@{\;}llc@{\;}}
\hline\hline
SF & Island Parameter & Range \\
\hline%
Aegean    & \texttt{floodclip}  & [2,5]\\
Caesar    & \texttt{mergeThr}   & [2,3]\\ 
ProFound  & \texttt{tolerance}  & [2,5]\\ 
PyBDSF    & \texttt{thresh\_isl} & [2,5]\\ 
Selavy    & \texttt{growthCut}  & [2,5]\\ 
\hline\hline
\end{tabular}
\label{tb:island_ranges}
\end{table}

Typhon will also perform an initial RMS box optimisation before optimising the PRD, if it is configured to do so. This is of particular importance for extended objects or around bright sources \citep{mohan_2015}, especially for Gaussian-based extraction SFs such as Aegean, PyBDSF, and Selavy. Typhon uses Aegean's background/noise image generation tool, \textsc{bane} \citep{hancock_2018}, to search the RMS box size (\texttt{box\_size}) and step size (\texttt{step\_size}) parameter space, 
\begin{equation}
    \left.\begin{array}{c}
         \mbox{$\displaystyle3\le\frac{\mbox{box\_size}}{[4(\texttt{BMAJ}+\texttt{BMIN})/2]}\le6$}\\\\
        \mbox{$\displaystyle\frac{1}{4}\le\frac{\mbox{step\_size}}{\mbox{box\_size}}\le\frac{1}{2}$}\\
    \end{array}\right\}\,,\label{eq:rms_box_pars}
\end{equation}
for the lowest background level, $\mu$ \citep[\textit{c.f.},][]{riggi_2016}. The $4(\texttt{BMAJ}+\texttt{BMIN})/2$ factor represents the \textsc{bane} default box size,
where we assume a square box, for simplicity.  The limits $3$ and $6$ are consistent with the rule of thumb that the box size should be 10 to 20 times larger than the beam size \citep{riggi_2016}. The $1/4$ and $1/2$ bounds are used for providing a smoothly sliding box \citep{mohan_2015}.

The Typhon optimisation algorithm can be summarised as follows.
\begin{itemize}
    \item If the RMS box $\mu$-optimisation is desired:
    \begin{itemize}
        \item Minimise $\mu$ over a $6\times3$ box\_size by step\_size search grid, constrained by Equation~\ref{eq:rms_box_pars}
    \end{itemize}
    \item Select a centralised $n\times n$ image sample-cutout:
    \begin{itemize}
        \item Use an $n^2$-area rectangle, if non-square image
        \item Use the full area, if image is too small 
    \end{itemize}
    \item Determine the RMS parameter bounds (Equation~\ref{eq:rms_pars})
    \item For each SF:
    \begin{itemize}
        \item If applicable, set the RMS box parameters to the optimised values
        \item Extract the island parameter bounds from Configuration Management (\textit{re.} Table~\ref{tb:island_ranges})
        \item Optimise the PRD of the sample-cutout:
        \begin{itemize}
            \item Iterate the RMS parameter backwards from RMS$_{max}$
            \item At each RMS step, iterate the island parameter, such that, $\sigma_{island}<\sigma_{RMS}$, otherwise set $\sigma_{island}=0.999\,\sigma_{RMS}$
            \item For each RMS-island parameter pair compute the PRD
            \item Terminate iterations just before the PRD passes below 90\%
            \begin{itemize}
                \item If the PRD is always below 90\% choose the highest value.
            \end{itemize}
        \end{itemize}
    \end{itemize}
    \item Run the SFs on the full image using their optimised parameters
    \item Archive the results in a tarball
\end{itemize}
\noindent For our initial studies, we have chosen to set $n=2.5^\circ$ to provide a sufficiently large region of sky to ensure the finder parameters are not biased by small-scale structure in a given image. Also Aegean, PyBDSF, and Selavy are configured to use the RMS box optimisation inputs from \textsc{bane}, with Selavy only accepting the RMS box size. Appendix~\ref{ap:source_finder_notes} provides details of the SFs and their settings used herein.

For the purposes of placing the SFs on equal footing, we have chosen to restrict Aegean, Caesar and Selavy to single threaded mode, so as to keep the background/noise statistics consistent, at the cost of computational speed. In addition, we keep all of the internal parameters of all of the SFs fixed, instead of tweaking them by hand for each use case. Every effort has been made to keep each SF module as generic as possible.

\subsubsection{Hydra}
\label{sc:hydra_sw}

\begin{figure*}[hbt!]
\centering%
\includegraphics[width=0.75\textwidth]{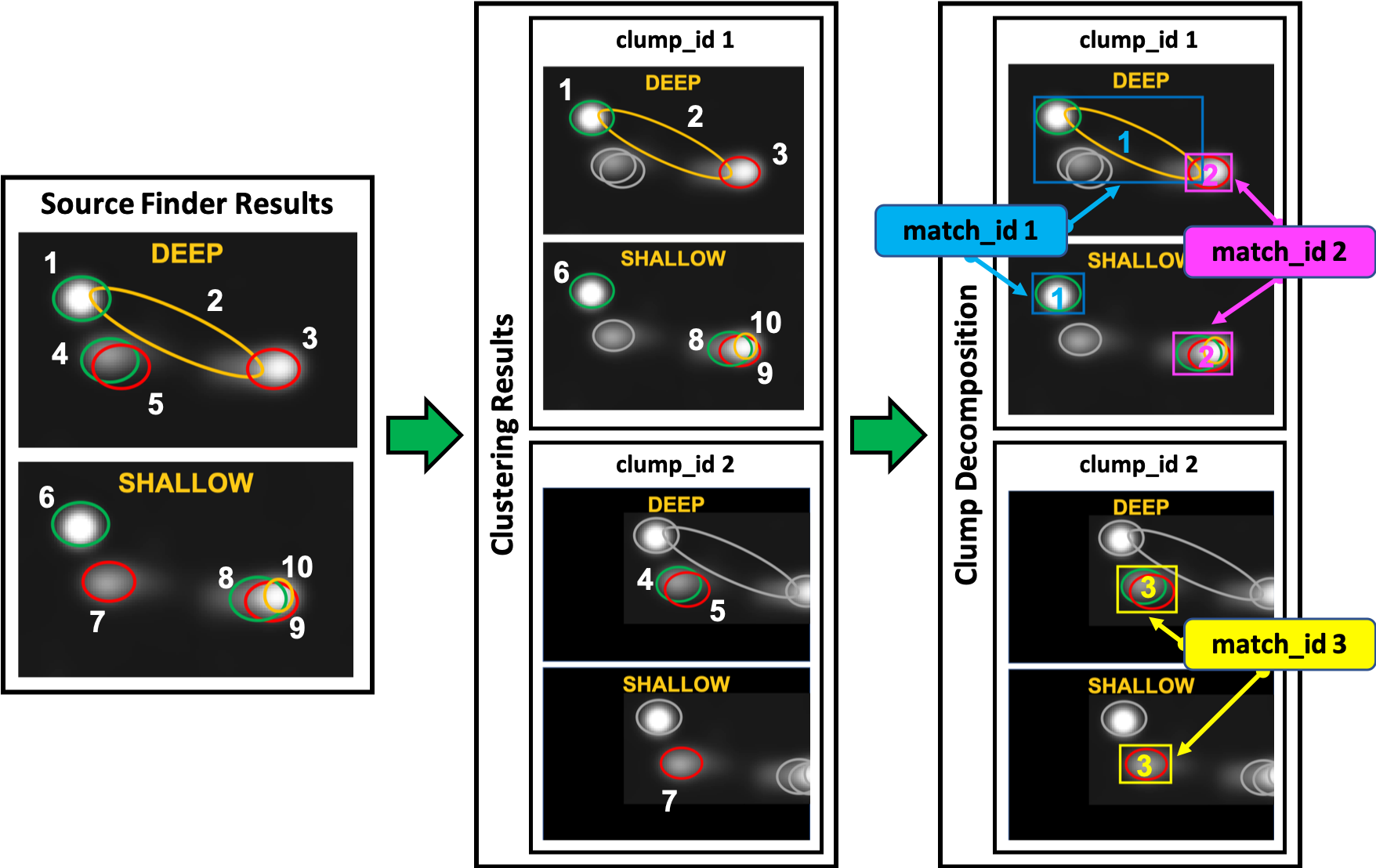}
\caption{\textbf{Clustering Algorithm Infographic:} The left panel shows the results of 3 hypothetical SFs (red, green, and gold). The middle panel shows the results after clustering, resulting in two clumps, assigned \texttt{clump\_id} 1 (upper panel) and \texttt{clump\_id} 2 (lower panel). A clump is defined through the spatial overlap between SF detections (\textit{i.e.}, components), in the $\mathcal{D}$ and $\mathcal{S}$-images together. The components are numbered independently and can be associated with the clump they end up in. For instance, components 1, 2, and 3 are linked together in the $\mathcal{D}$-image, and, in addition, 1 overlaps with 6, and 3 overlaps with 8, 9, and 10 in the $\mathcal{S}$-image. Together this set of components populate \texttt{clump\_id} 1. Similarly, \texttt{clump\_id} 2 is composed of components 4, 5, and 7. Clumps are centered in the Hydra Viewer (Figure~\ref{fg:hydra_viewer}), with unassociated components greyed out. The right panel shows the results after the clumps are decomposed into closest (\textit{i.e.}, overlapping center-to-center) matches between SFs, in the $\mathcal{D}$ and $\mathcal{S}$-images, such that there is only one SF with a match in the $\mathcal{D}$ and $\mathcal{S}$-images. These matched sets are assigned \texttt{match\_id}s, with boxes enclosing the extremities of the components. The Hydra Viewer displays these numbers at the center of the boxes (which are coloured differently here, for emphasis). So \texttt{clump\_id} 1 contains \texttt{match\_id} 1 = \{1, 2, 6\} and \texttt{match\_id} 2 = \{3, 8, 9, 10\}, while \texttt{clump\_id} 2 contains \texttt{match\_id} 3 = \{4, 5, 7\}.}
\label{fg:clustering_infofographic}
\end{figure*}

\begin{figure*}[htb!]
\begin{center}
\includegraphics[width=0.6\textwidth]{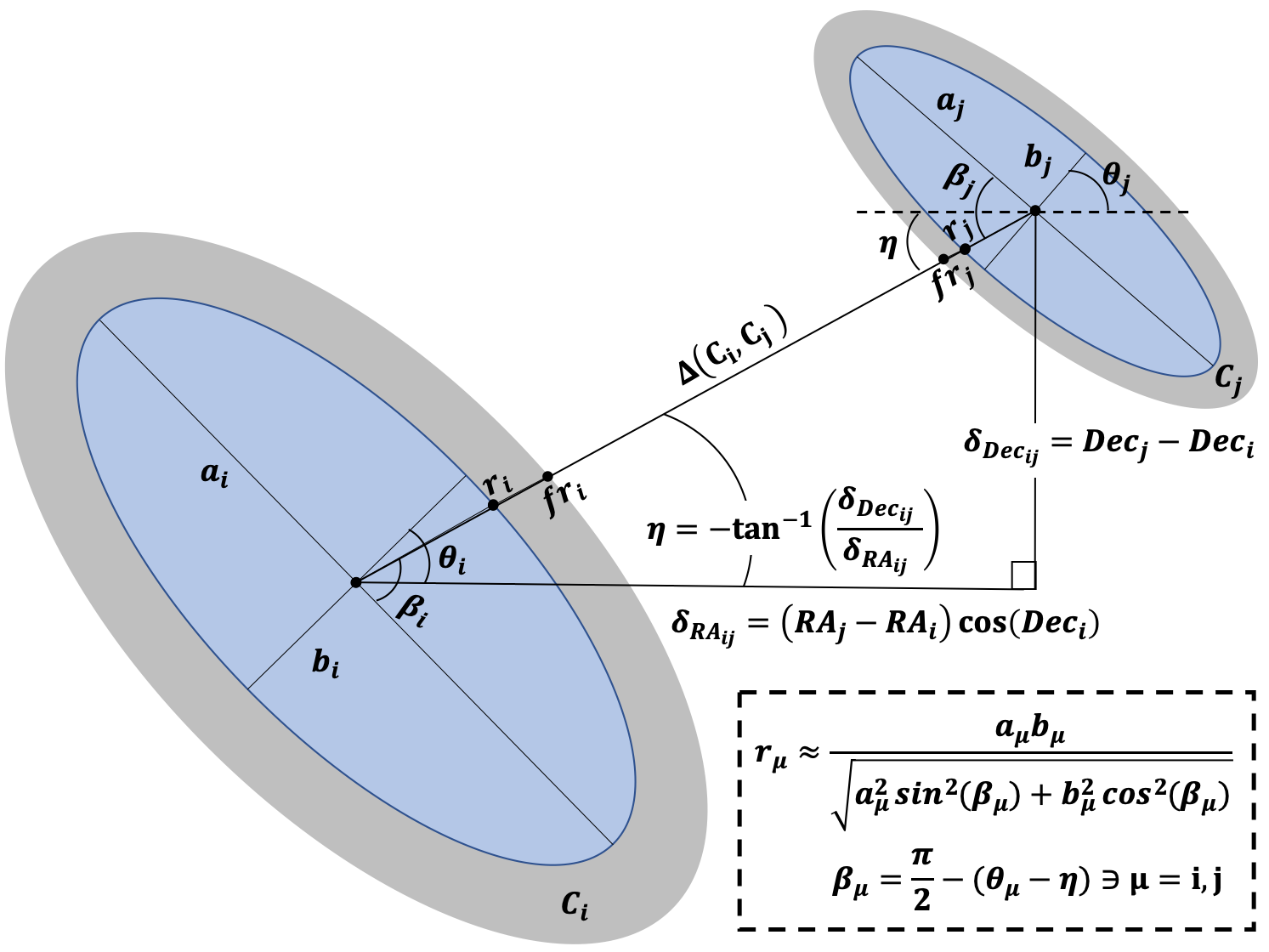}
\caption{Derivation of the distance metric used for clustering. Here we assume that the space is locally flat, so that  $\Delta(C_i,C_j)\approx(\delta^2_{\texttt{RA}_{ij}}+\delta^2_{\texttt{Dec}_{ij}})^{1/2}$, where $\delta_{\texttt{RA}_{ij}}=(\texttt{RA}_j-\texttt{RA}_i)\cos(\texttt{Dec}_i)$ and $\delta_{\texttt{Dec}_{ij}}=\texttt{Dec}_j-\texttt{Dec}_i$. The distances from the centers of components $C_i$ and $C_j$ to their edges, along a ray between them, is given by $r_i$ and $r_j$, respectively: \textit{i.e.}, $r_\mu$, is a standard geometrical expression in terms of angle $\beta_\mu=\pi/2-(\theta_\mu-\eta)$ with respect to the ray and the semi-major axis $a_\mu$, where $\theta_\mu$ is the position angle, $\eta=-\tan^{-1}(\delta_{\texttt{Dec}_{ij}}/\delta_{\texttt{RA}_{ij}})$, and $\mu=i,\,j$. The grey area outside the ellipses is the skirt, whose extent is determined by $f$.
\label{fg:clustering_metric}}
\end{center}
\end{figure*}

Hydra is the main tool that glues everything together, by running Typhon for $\mathcal{D}$ and $\mathcal{S}$ images, and producing data products such as diagnostics plots and catalogues. One of main underlying features of Hydra is that it uses a clustering algorithm \citep{boyce_2018} to relate information between SF components in both $\mathcal{D}$ and $\mathcal{S}$ images. In addition, Hydra will also accept simulated catalogue input, under a source-finder pseudonym.

Figure~\ref{fg:clustering_infofographic} shows an example of how the clustering algorithm works. All components between all $\mathcal{D}$ and $\mathcal{S}$ SF detections (\textit{i.e.}, catalogue rows) are spatially grouped together, with their overlaps forming clumps with unique \texttt{clump\_id}s. The clumps are also decomposed into the closest $\mathcal{DS}$ matches, and assigned unique \texttt{match\_id}s. The matches are further broken down by SF into subclumps (not shown), and assigned unique \texttt{subclump\_id}s. All of this information is compiled into a cluster catalogue (or table), containing the following key reference elements (columns): cluster catalogue ID, clump ID, match ID, subclump ID, SF $\mathcal{D/S}$ catalogue cross-reference ID, and image depth ($=\mathcal{D},\,\mathcal{S}$). In addition, the catalogue contains common SF output parameters, such as RA, Dec, flux density, \textit{etc.} There is also a clump catalogue, consisting of rows by unique \texttt{clump\_id}, of cluster centroid positions, cutout sizes, total number of components, number of components per SF, SFs with the best residual statistics, \textit{etc.}

Figure~\ref{fg:clustering_metric} shows the derivation of the distance metric used in the clustering algorithm. The algorithm uses the following distance metric constraint to determine the overlap between two elliptical components, $C_i$ and $C_j$, with center-to-edge distances, $r_i$ and $r_j$, along an adjoining ray.
\begin{equation}
    \Delta(C_i,C_j)\le r_i+r_j\label{eq:clustering_metric}
\end{equation}
\noindent where
\begin{equation}
    \Delta(C_i,C_j)=\sqrt{(\texttt{RA}_j\!-\!\texttt{RA}_i)^2\cos^2(\texttt{Dec}_i)\!+\!(\texttt{Dec}_j\!-\!\texttt{Dec}_i)^2}\,,\nonumber
\end{equation}
is the distance metric, and
\begin{equation}
    r_\mu=\frac{a_\mu b_\mu}{\sqrt{a_\mu^2\cos^2(\theta_\mu-\eta)+b_\mu^2\sin^2(\theta_\mu-\eta)}}\,,\nonumber
\end{equation}
\noindent are the center-to-edge distances, for $\mu=i,j$, with
\begin{equation}
    \eta = -\tan^{-1}\left[\frac{\texttt{Dec}_j-\texttt{Dec}_i}{(\texttt{RA}_j-\texttt{RA}_i)\cos(\texttt{Dec}_i)}\right]\,,\nonumber
\end{equation}
$a_\mu$ is the semi-major axis, $b_\mu$ is the semi-minor axis, and $\theta_\mu$ is the position angle \citep[defined in the same manner as the beam \texttt{PA},][]{greisen_2017}. So components satisfying this constraint are clumped together.

\begin{figure*}[htb!]
\begin{center}
\includegraphics[width=\textwidth]{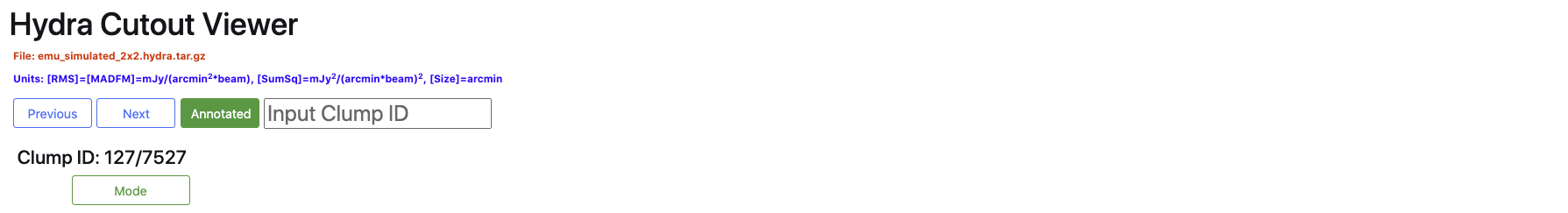}\\
\includegraphics[width=\textwidth]{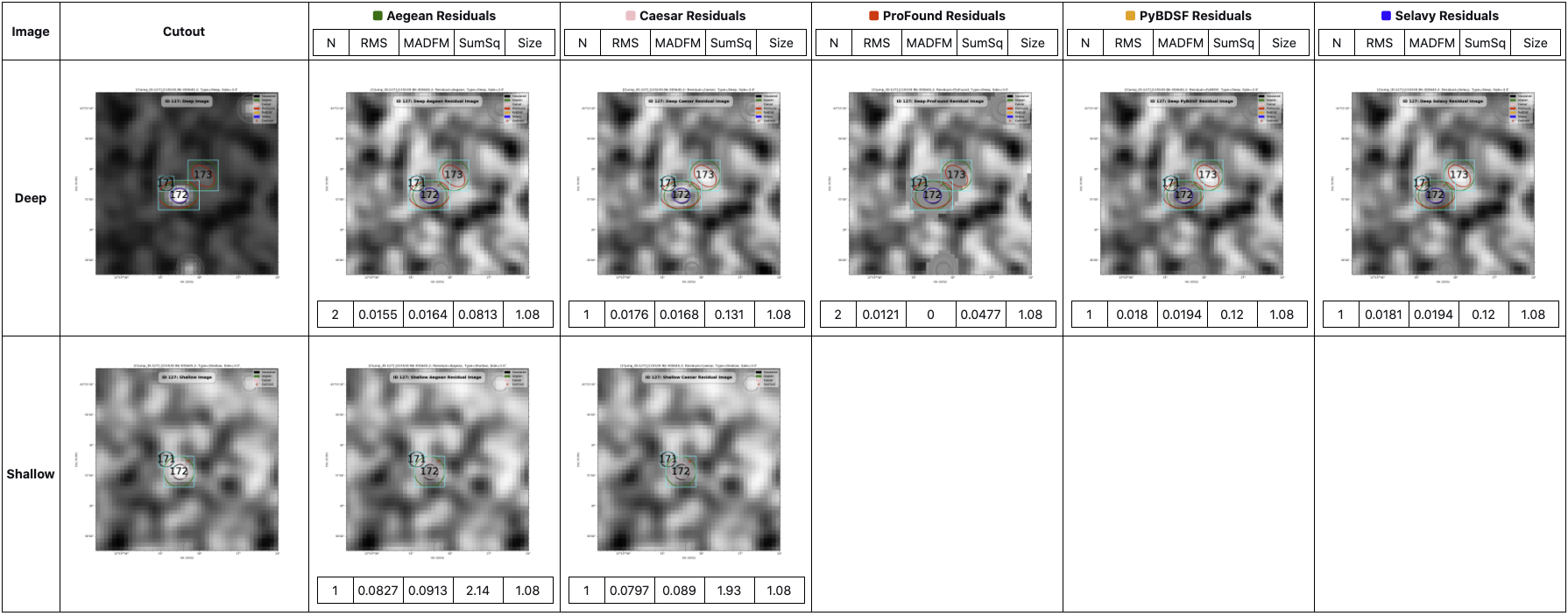}\\
\includegraphics[width=\textwidth]{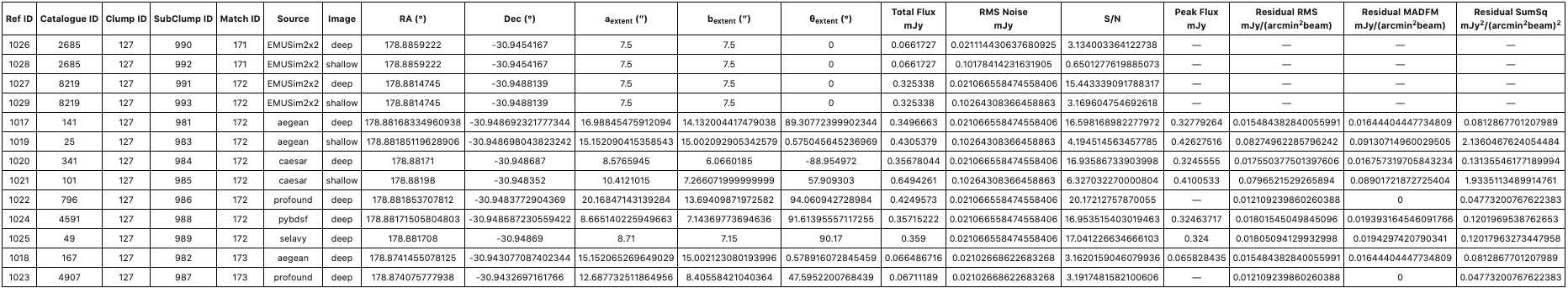}
\caption{\textbf{Hydra Viewer Infographic:} The cutout viewing section of Hydra's local web-viewer interface. At the top is the navigation bar, which allows the user to navigate by clump ID, go to a specific clump, turn on/off cutout annotations, or examine S/N bins of diagnostic plots such as $\mathcal{C}$ and $\mathcal{R}$, by using the Mode button. The main panel contains $\mathcal{D}$ (top) and $\mathcal{S}$ (bottom) square image cutouts (first column) and SF residual image cutouts (remaining columns), centered about a given clump's centroid. Here the annotation is turned on, with the neighboring clumps greyed out. The table at the bottom (not to scale) is the cluster table rows for the clump, with the following columns: cluster catalogue ID, SF catalogue cross-reference ID, clump ID, subclump ID, match ID, SF or $\mathcal{J}$ catalogue name, image depth, RA ($^\circ$), Dec ($^\circ$), semi-major axis ($^{\prime\prime}$), semi-minor axis  ($^{\prime\prime}$), position angle ($^\circ$), total flux-density (mJy), \textsc{bane} RMS noise (mJy), S/N (total flux over \textsc{bane} RMS noise), peak flux (mJy/beam), normalized-residual RMS (mJy/(arcmin$^2$ beam)), normalized-residual MADFM (mJy/(arcmin$^2$ beam)), and normalized-residual $\Sigma I^2$ ((mJy/(arcmin$^2$ beam))$^2$).  The normalized-residual statistics are normalised by the cutout area (arcmin$^2$). This statistical information is also shown below each cutout, along with the number of components (N), and cutout size (Size, in arcmin). This figure is to illustrate the layout of the Hydra viewer, not the details. It shows screen shots from the Hydra Viewer pasted together, hence the fonts appear small. The data at the bottom is raw output from the cluster table, which is not rounded in this version of the software.
\label{fg:hydra_viewer}}
\end{center}
\end{figure*}

\begin{table}[hbt!]
\caption{SF annotation colours.}
\label{tb:annotation_colours}
\centering
\begin{tabular}{@{\;}ll@{\;}}
\hline\hline
SF & Colour \\
\hline
Aegean & Green \\
Caesar & Magenta \\
ProFound & Red \\
PyBDSF & Orange \\
Selavy & Blue \\
Simulated & Black \\
\hline\hline
\end{tabular}
\end{table}

\begin{figure}[htb!]
\begin{center}
\includegraphics[width=\columnwidth]{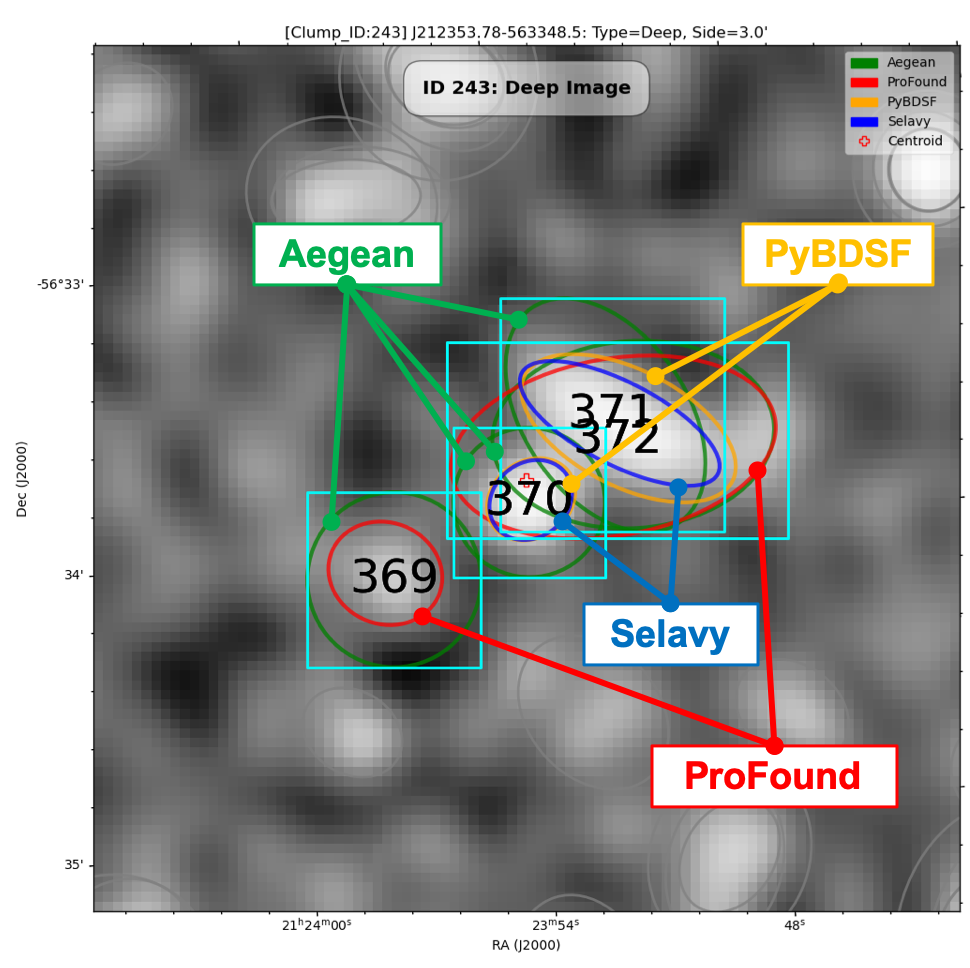}
\caption{An example of a $\mathcal{D}$-image cutout, with annotations turned on, consisting of 4 Aegean, 2 ProFound, 2 PyBDSF, and 2 Selavy overlapping $\mathcal{D}$-image catalogue components. The label at the top indicates it corresponds to \texttt{clump\_id} 243, and the numbers at the centers of the cyan boxes are the \texttt{match\_id}s (369 through 372). \label{fg:pilot_clump_id243_example}}
\end{center}
\end{figure}

Hydra also provides a web-viewer (known as the Hydra Viewer) for exploring image and residual-image cutouts by \texttt{clump\_id}, along with corresponding cluster table information. Figure~\ref{fg:hydra_viewer} provides a detailed description of the Hydra Viewer's cutout interface.
As indicated in the figure, the viewer has radio component annotations that can be toggled on/off. Figure~\ref{fg:pilot_clump_id243_example} provides a more detailed example. Table~\ref{tb:annotation_colours} describes the annotation colours, which are stored as metadata in Hydra's Configuration Management (Figure~\ref{fg:Cerberus_Code_Generation}).

The following is a list of the data products produced by Hydra:
\begin{itemize}
    \item Typhon Metrics
    \begin{itemize}
        \item $\mathcal{D/S}$ Diagnostic Plots of
        \begin{itemize}
            \item PRD
            \item PRD CPU Times
            \item Residual RMS
            \item Residual MADFM
            \item Residual $\Sigma I^2$
        \end{itemize}
        \item Table of $\mathcal{D}$ and $\mathcal{S}$ optimised RMS and island parameters
    \end{itemize}
    \item $\mathcal{D/S}$ Catalogues
    \begin{itemize}
        \item SF Catalogues
        \item Cluster Catalogue
        \item Clump Catalogue
    \end{itemize}
    \item Optional Simulated Input Catalogue
    \item $\mathcal{D/S}$ Cutouts
    \begin{itemize}
        \item Un/annotated Images
        \item Un/annotated Residual Images
    \end{itemize}
    \item $\mathcal{D/S}$ Diagnostic Plots
    \begin{itemize}
        \item Clump Size Distributions
        \item Detections \textit{vs.} S/N
        \item $\mathcal{C}$ \textit{vs.} S/N
        \item $\mathcal{R}$ \textit{vs.} S/N
        \item Flux-Ratios: $S_{out}/S_{in}$ \textit{vs.} S/N
        \item False-Positives  \textit{vs.} S/N
    \end{itemize}
    \item Hydra Viewer: Local Web-browser Tool
\end{itemize}
\noindent All of this information is stored in a tarball. The Hydra Viewer allows the user to view all of these data products. The cutout viewer portion is linked only to the cluster catalogue. It is accessible through an \texttt{index.html} file in the main \texttt{tar} directory.

\section{Completeness and Reliability Metrics}
\label{sc:oh_goody}
\begin{figure}[htb!]
    \centering
    \includegraphics[width=\columnwidth]{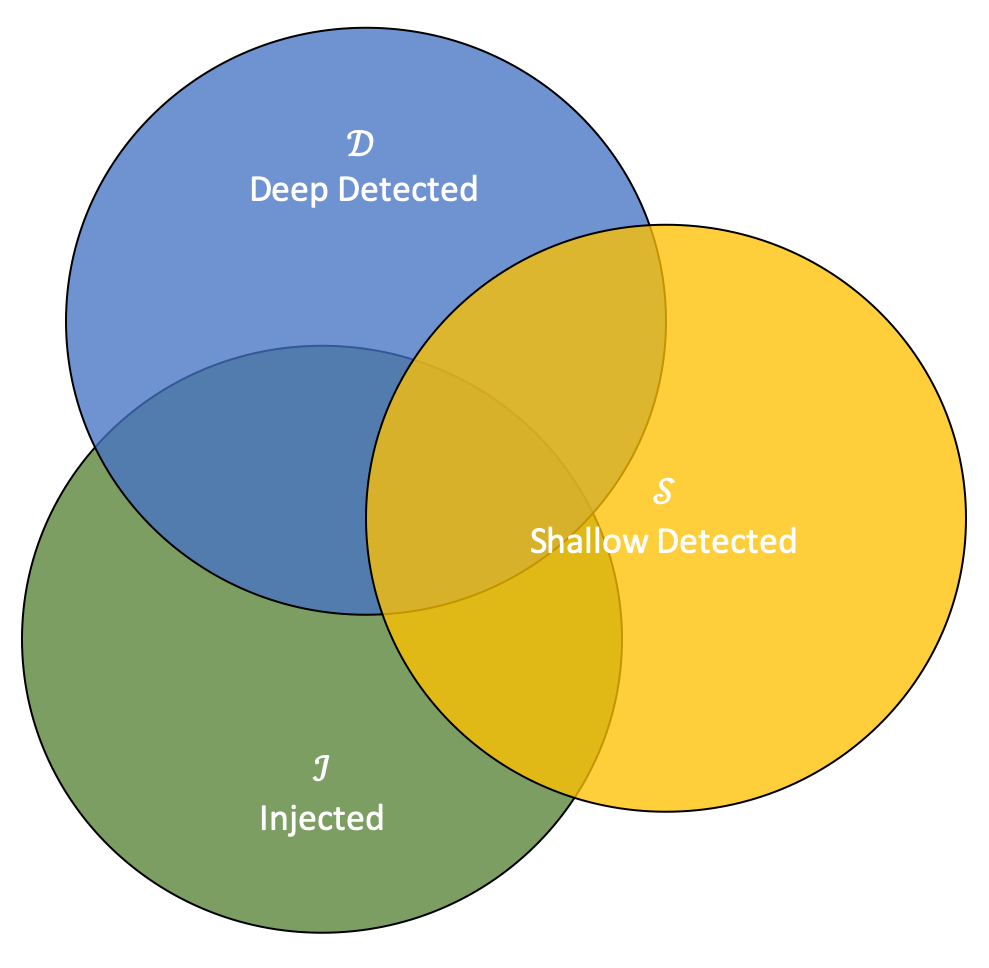}
    \caption{Venn diagram of completeness and reliability, for sets of deep ($\mathcal{D}$), shallow ($\mathcal{S}$), and injected ($\mathcal{J}$) sources.}
    \label{fg:cr_venn_diagram}
\end{figure}

Completeness and reliability metrics can be generated through various combinations of deep, shallow, and injected sources. Figure~\ref{fg:cr_venn_diagram} shows a Venn diagram of the overlapping possibilities. In addition, we need to be careful in our definitions of these metrics. 

\label{ap:completeness_and_reliability}

Here we use a clustering approach to spatially match our detections (Equation~\ref{eq:clustering_metric}). An alternative method is to use a cutoff distance in a catalogue cross-match \citep[\textit{e.g.},][]{huynh_2012,hopkins_2015,hale_2019}.\footnote{\cite{bonaldi_2021} also requires a level of consistency in flux density to associate independently detected sources as being in common. We do not use this in our metric as it can lead to falsely rejecting true  source associations where flux densities are poorly measured.} Depending on the cutoff distance adopted, ``locally,'' this approach may lead to associations between adjacent clumps that may not actually be related. The clustering approach aims to mitigate against this effect.

Completeness ($\mathcal{C}$) is the fraction of real detections to real sources, and reliability ($\mathcal{R}$) is the fraction of real detections to detected sources (Figure~\ref{fg:cr_venn_diagram}). Here we define these metrics in terms of ``real'' injected (simulated) sources ($\mathcal{J}$) \textit{vs.} deep ($\mathcal{D}$) and shallow ($\mathcal{S}$) detections and, ``assumed-real'' deep detections \textit{vs.} shallow detections. In the case where the sources are known (\textit{i.e.}, injected), we take the fraction of real--deep ($\mathcal{D}\cap\mathcal{J}$) or real--shallow ($\mathcal{S}\cap\mathcal{J}$) detections to the injected sources for our completeness,
\begin{equation}
    \mathcal{C_{\mathcal{D}}}=\frac{\mathcal{D}\cap\mathcal{J}}{\mathcal{J}}\label{eq:completeness_deep}
\end{equation}
\noindent or 
\begin{equation}
    \mathcal{C_{\mathcal{S}}}=\frac{\mathcal{S}\cap\mathcal{J}}{\mathcal{J}}\,,\label{eq:completeness_shallow}
\end{equation}
\noindent respectively.\footnote{In general, in our notation, the length of a set is implicitly assumed: \textit{e.g.}, $\mathcal{D}\cap\mathcal{J}\equiv|\mathcal{D}\cap\mathcal{J}|$.}  Similarly, the fraction of real--deep or real--shallow detections to the corresponding deep or shallow detections give the reliability
\begin{equation}
    \mathcal{R_{\mathcal{D}}}=\frac{\mathcal{D}\cap\mathcal{J}}{\mathcal{D}}\label{eq:reliability_deep}
\end{equation}
\noindent or
\begin{equation}
    \mathcal{R_{\mathcal{S}}}=\frac{\mathcal{S}\cap\mathcal{J}}{\mathcal{S}}\,,\label{eq:reliability_shallow}
\end{equation}
\noindent respectively. In the case where no true underlying sources are known we use the deep detections as a proxy, and take the fraction of real--shallow detections to deep detections
for our completeness,
\begin{equation}
    \mathcal{C_{\mathcal{DS}}}=\frac{\mathcal{S}\cap\mathcal{D}}{\mathcal{D}}\,,\label{eq:completeness_deep_shallow}
\end{equation}
\noindent and the fraction of real--shallow to shallow detections for our reliability,
\begin{equation}
    \mathcal{R_{\mathcal{DS}}}=\frac{\mathcal{S}\cap\mathcal{D}}{\mathcal{S}}\,.\label{eq:reliability_deep_shallow}
\end{equation}

We can take this one step further by asking the question, \textit{``Given our knowledge of injected sources, how good are our measures of deep--shallow completeness ($\mathcal{C_{\mathcal{DS}}}$) and reliability ($\mathcal{R_{\mathcal{DS}}}$)?''} From this, we define the fraction of real--deep--shallow detections,  $(\mathcal{D}\cap\mathcal{J})\cap(\mathcal{S}\cap\mathcal{J})$, to real--deep detections, $\mathcal{D}\cap\mathcal{J}$, as our goodness of completeness,
\begin{equation}
    \tilde{C}_{\mathcal{DS}}=\frac{(\mathcal{D}\cap\mathcal{J})\cap(\mathcal{S}\cap\mathcal{J})}{\mathcal{D}\cap\mathcal{J}}\,,\label{eq:godness_of_completeness}
\end{equation}
\noindent and the fraction of real--deep--shallow detections to real-shallow detections, $\mathcal{S}\cap\mathcal{J}$, as our goodness of reliability,
\begin{equation}
    \tilde{R}_{\mathcal{DS}}=\frac{(\mathcal{D}\cap\mathcal{J})\cap(\mathcal{S}\cap\mathcal{J})}{\mathcal{S}\cap\mathcal{J}}\,.\label{eq:godness_of_reliability}
\end{equation}
\noindent Table~\ref{tb:cr_metrics} summarises all of our completeness and reliability metrics (Equations~\ref{eq:completeness_deep} through~\ref{eq:godness_of_reliability}).\footnote{The current version of the Hydra Viewer (Figure~\ref{fg:hydra_viewer}) only supports filtering the S/N bins of its $\mathcal{C_{\mathcal{DS}}}$ and $\mathcal{R_{\mathcal{DS}}}$ diagnostic plots, through its Mode button.}

\begin{table*}[htb!]
\caption{Completeness/reliability metrics (see Figure~\ref{fg:cr_venn_diagram}) in terms of deep ($\mathcal{D}$), shallow ($\mathcal{S}$), and injected ($\mathcal{J}$) sources.}
\centering
\begin{tabular}{@{\;}lllll@{\;}}
\hline\hline
Inputs & Detections & Real Detections & Completeness & Reliability \\
\hline%
&&&&\\
$\mathcal{J}$ & $\mathcal{D}$ & $\mathcal{D}\cap\mathcal{J}$
   & ${\displaystyle\mathcal{C_{\mathcal{D}}}=\frac{\mathcal{D}\cap\mathcal{J}}{\mathcal{J}}}$ 
   & ${\displaystyle\mathcal{R_{\mathcal{D}}}=\frac{\mathcal{D}\cap\mathcal{J}}{\mathcal{D}}}$\\ 
&&&&\\
$\mathcal{J}$ & $\mathcal{S}$ & $\mathcal{S}\cap\mathcal{J}$ 
   & ${\displaystyle\mathcal{C_{\mathcal{S}}}=\frac{\mathcal{S}\cap\mathcal{J}}{\mathcal{J}}}$ 
   & ${\displaystyle\mathcal{R_{\mathcal{S}}}=\frac{\mathcal{S}\cap\mathcal{J}}{\mathcal{S}}}$\\ 
&&&&\\
$\mathcal{D}$ & $\mathcal{S}$ & $\mathcal{S}\cap\mathcal{D}$
   & ${\displaystyle\mathcal{C_{\mathcal{DS}}}=\frac{\mathcal{S}\cap\mathcal{D}}{\mathcal{D}}}$ 
   & ${\displaystyle\mathcal{R_{\mathcal{DS}}}=\frac{\mathcal{S}\cap\mathcal{D}}{\mathcal{S}}}$\\ 
&&&&\\
$\mathcal{D}\cap\mathcal{J}$ & $\mathcal{S}\cap\mathcal{J}$ & $(\mathcal{D}\cap\mathcal{J})\cap(\mathcal{S}\cap\mathcal{J})$
   & ${\displaystyle\tilde{C}_{\mathcal{DS}}=\frac{(\mathcal{D}\cap\mathcal{J})\cap(\mathcal{S}\cap\mathcal{J})}{\mathcal{D}\cap\mathcal{J}}}$ 
   & ${\displaystyle\tilde{R}_{\mathcal{DS}}=\frac{(\mathcal{D}\cap\mathcal{J})\cap(\mathcal{S}\cap\mathcal{J})}{\mathcal{S}\cap\mathcal{J}}}$\\ 
&&&&\\
\hline\hline
\end{tabular}
\label{tb:cr_metrics}
\end{table*}

\begin{figure}[htb!]
    \centering
    \includegraphics[width=\columnwidth]{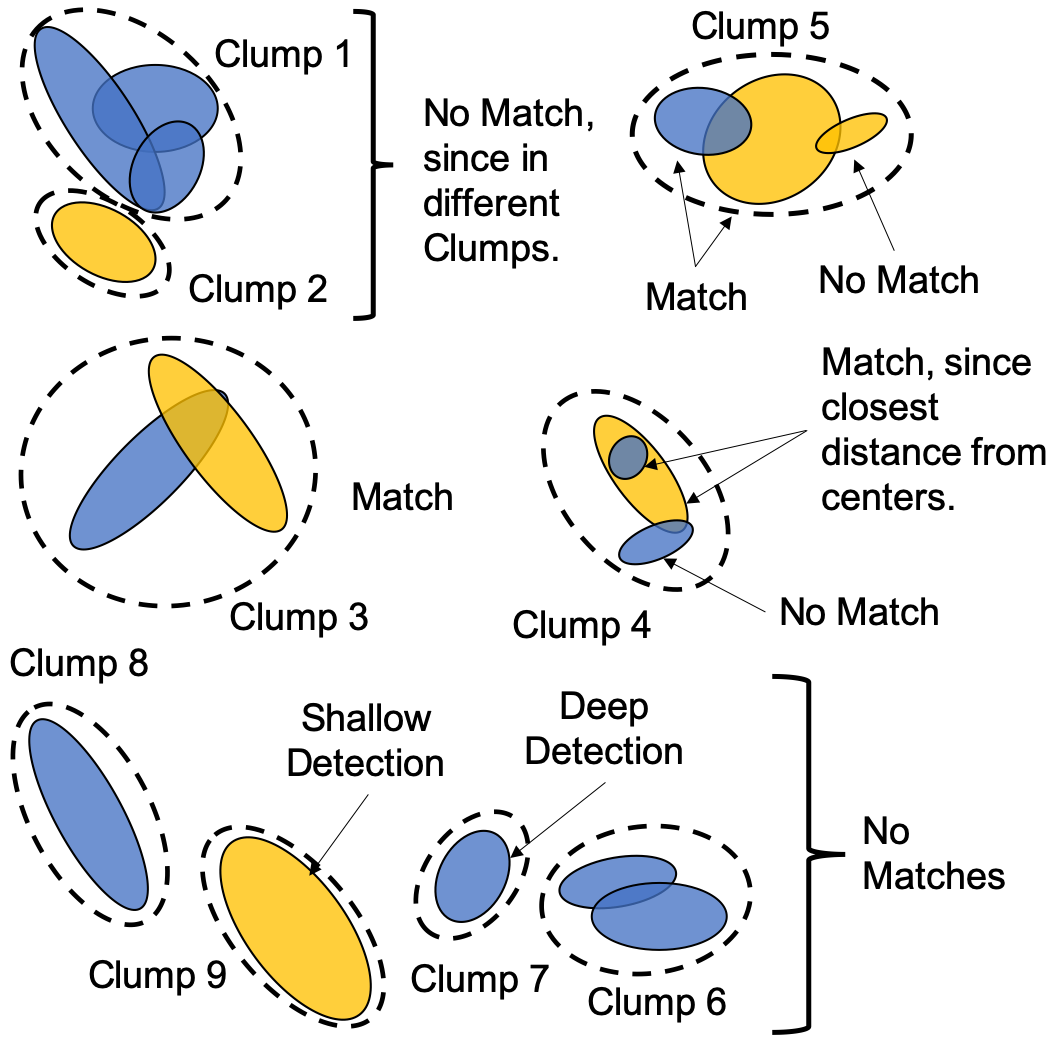}
    \caption{Examples of deep (blue) and shallow (amber) source component overlaps, $\mathcal{C_{\mathcal{DS}}}=(\mathcal{S}\cap\mathcal{D})/\mathcal{D}$ and $\mathcal{R_{\mathcal{DS}}}=(\mathcal{S}\cap\mathcal{D})/\mathcal{S}$. Real-shallow detections are indicated by overlapping pair-wise deep-shallow detections ($\mathcal{S}\cap\mathcal{D}$), whose centers are closest. The dash-lines indicate clumps of component extent overlays.}
    \label{fg:ds_overlaps}
\end{figure}

Figure~\ref{fg:ds_overlaps} shows examples of deep--shallow source component overlaps, $\mathcal{S}\cap\mathcal{D}$, to illustrate the calculation of $\mathcal{C_{\mathcal{DS}}}$ and $\mathcal{R_{\mathcal{DS}}}$. Matches are done pair-wise, within clumps, between the closest centers of overlapping deep--shallow components. This method is more precise than a typical fixed separation nearest neighbour approach \citep{hopkins_2015,riggi_2019}, as it ensures source components always overlap. The  $|\mathcal{S}\cap\mathcal{D}|:|\mathcal{D}|$ and $|\mathcal{S}\cap\mathcal{D}|:|\mathcal{S}|$ ratios are then binned with respect to the $\mathcal{D}$ and $\mathcal{S}$ flux densities (or S/N), respectively, producing $\mathcal{C_{\mathcal{DS}}}$ \textit{vs.} $\mathcal{D}$ completeness and $\mathcal{R_{\mathcal{DS}}}$ \textit{vs.} $\mathcal{S}$ reliability histograms.

\section{Validation}
\label{sc:validation}
In this section we use $2^\circ\times2^\circ$ simulated-compact (CMP) and simulated-extended (EXT) image data to characterise the performance of Hydra, and validate some new metrics. In particular, the simulated data is used to explore and develop metrics that can be used for real images where the ground-truth is unknown. A preliminary discussion on SF performance is also presented. Paper~II is focused on cross-SF comparison, using our simulated data along with some real data.

\subsection{Image Data}
\label{sc:image_data}
\subsubsection{Simulated Compact Sources, CMP}
\label{sc:sim_point_sources}
\begin{figure}[hbt!]
\centering%
\includegraphics[scale=0.3]{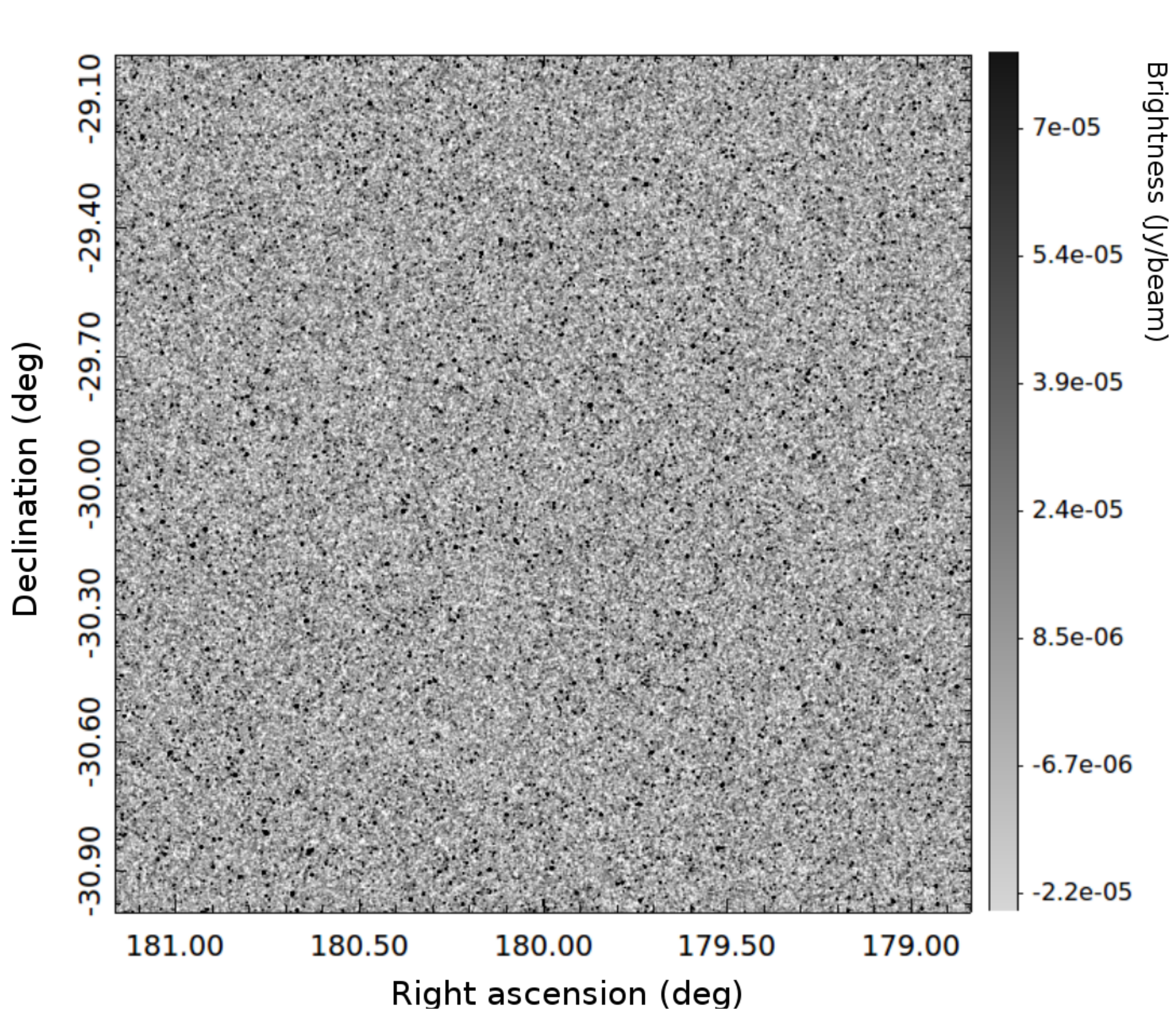}
\caption{Simulated map with point-like (compact) sources. The coordinates are arbitrarily set, and the FWHM is set to $15''$.}
\label{fg:simulated_point_sources}
\end{figure}

The simulated image, shown in Figure~\ref{fg:simulated_point_sources}, is produced in two steps; generation of a noise image, followed by the addition of artificial sources. We use {\sc miriad} \citep{sault_1995} to generate the artificial noise image, using the following steps. The {\sc imgen} task was used to produce a 1800$\times$1800 pixel image, with $4''$ pixels, (\textit{i.e.}, a $2^{\circ} \times 2^{\circ}$ field) populated by random Gaussian noise of RMS $20\,\mu$Jy/beam. This image was convolved using {\sc convol} to mimic a $15''$ FWHM beam, which has the effect of increasing the noise level by a factor of 2.8, so the resulting image is then scaled using {\sc maths} to divide by this factor, restoring the original noise level of $20\,\mu$Jy/beam.

To generate the properties of the artificial sources, we use the 6th order polynomial fit to the 1.4\,GHz source counts from \citet{hopkins_2003}, which is consistent with more recent source count determinations for the flux density range considered here \citep[e.g.,][and references therein]{gurkan22}. A sequence of 34 exponentially spaced bins in flux density was defined, ranging from $50\,\mu$Jy to $1\,$Jy, and within each bin the number of sources was calculated from the source count model. The flux density for each artificial source was assigned randomly between the extrema of the bin in which it lies. Source positions were also assigned randomly, with no attempt to mimic the clustering properties of real sources. For the $2^{\circ} \times 2^{\circ}$ field with a flux density limit of $50\,\mu$Jy, we end up with a list of 9,075 artificial sources.

These sources were added to the noise image using the Python module Astropy \citep{astropy_2013, astropy_2018} by constructing 2D Gaussian models with the FWHM of the restoring beam (15$''$) and scaling the amplitude to represent the randomly assigned peak flux density of the source. Given the sources modelled here are assumed to be point-like (compact), the peak flux density for a source has the same amplitude as the integrated flux density. Using this Gaussian model for each source, we generated an image array for each source to be added into the simulated image. We used Astropy again to convert the RA/Dec location of the source to pixel locations and each source was added to the simulated image at the desired location. 

\subsubsection{Simulated Extended Sources, EXT}
\label{sc:sim_extended_sources}
Following a similar procedure as in Section~\ref{sc:sim_point_sources}, we generated a sky model of size 1800$\times$1800 pixels ($4^{\prime\prime}$ pixel size, 2$^{\circ}\times$2$^{\circ}$ field of view) with both point-like and extended sources injected. The noise level is again set to 20\,$\mu$Jy/beam. Extended sources are 2D elliptical Gaussians with randomised position angle and axis ratio, with axis ratio varying between 0.4 to 1. A maximum major axis size was set at three times the synthesised beam size ($45^{\prime\prime}$, with a $15^{\prime\prime}$ FWHM beam, as in Section ~\ref{sc:sim_point_sources}). 

\begin{figure}[hbt!]
\centering%
\includegraphics[scale=0.3]{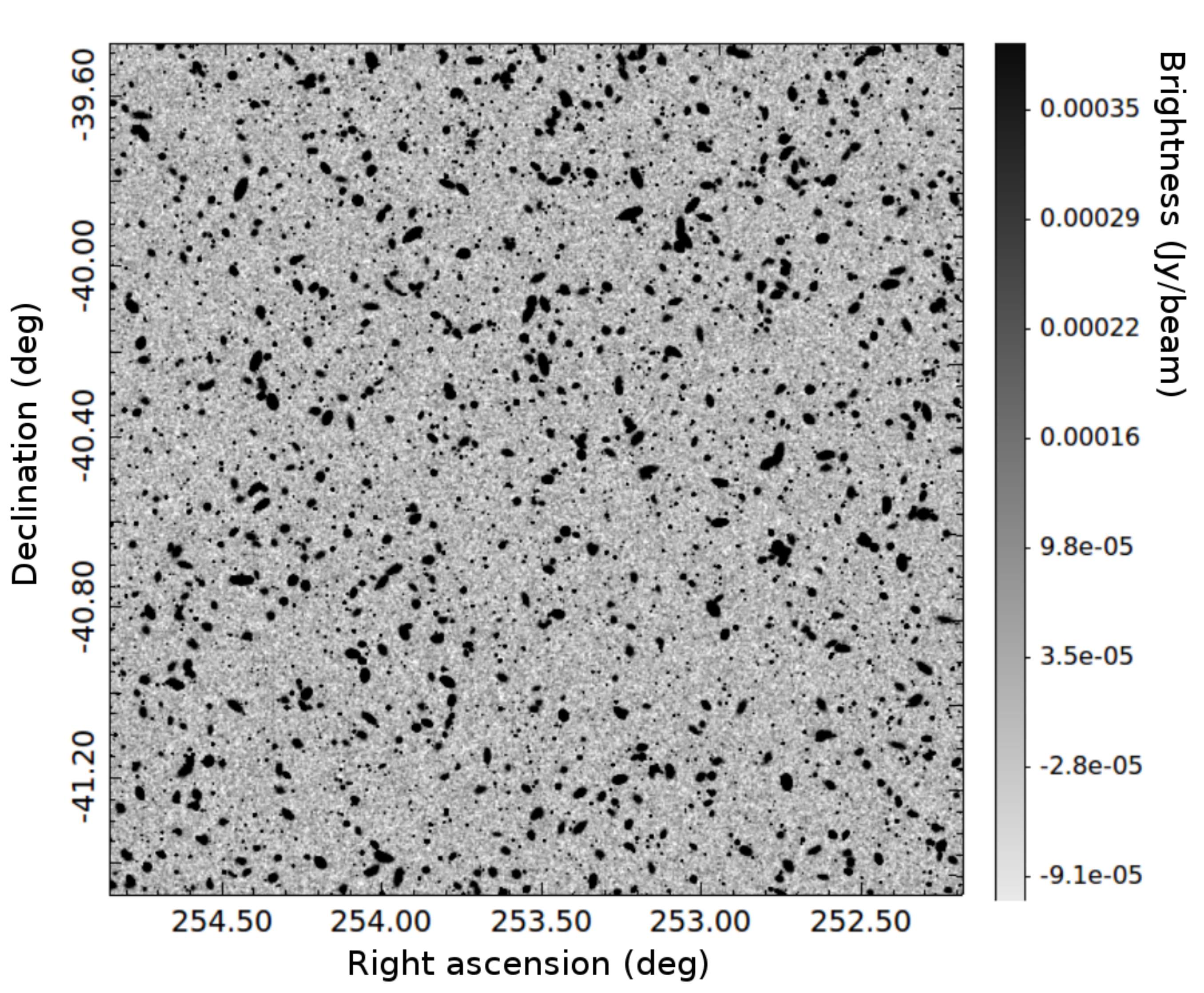}
\caption{Simulated image with both point-like (compact) and extended sources. The sky coordinates are arbitrarily chosen.}
\label{fg:extended_source_sim}
\end{figure}

A total of 9,974 artificial sources were injected, corresponding to a source density of 2,500 deg$^{-2}$, with 10\% being extended sources.
The peak flux densities $S$ of both point-like and extended sources were set to follow an exponential distribution $10^{-\lambda S}$ with $\lambda$=1.6, consistent with that seen in early ASKAP observations of the Stellar Continuum Originating from Radio Physics In our Galaxy \citep[\textsc{Scorpio},][]{umana_2015} field \citep{riggi_2021}. The minimum peak flux density for all sources was set at $50\,\mu$Jy and the maximum fixed at 1\,Jy for compact sources and 1\,mJy for extended sources. The final simulated image, shown in Figure~\ref{fg:extended_source_sim}, was produced by convolving this input sky model using the CASA\footnote{Common Astronomy Software Applications, \url{https://casa.nrao.edu/}} \textit{imsmooth} task and a target resolution of $15^{\prime\prime}$.

It is important to note here that, unlike the compact source simulation above where the faintest injected source lies at $\mbox{(S/N)}_{min}\sim2.5$, in the extended source simulated image 20.6\% of the injected sources fall below $\mbox{S/N}=1$. This is by design, to provide a more realistic image, and to test the impact on the SFs of having real sources lying below the noise level \citep[\textit{re.}][]{boyce_2022b}.

\subsection{Typhon Statistics}
\label{sc:typhon_stats}
\subsubsection{Optimisation Run Results}
Hydra uses Typhon to set the RMS box and island parameters for each SF (Aegean, Caesar, ProFound, PyBDSF, and Selavy) to ensure a 90\% PRD cutoff. The RMS box parameters, obtained from Typhon's $\mu$-optimization routine, were used by Aegean, PyBDSF, and Selavy. An $\mathcal{S}$-image was generated by Homados, and RMS box and island parameters similarly estimated.
Tables~\ref{tb:hydra_2x2_typhon_rms_box_stats} and~\ref{tb:hydra_2x2_typhon_stats} summarise these results for our CMP and EXT images.

\begin{table}[hbt!]
\caption{Hydra $\mu$-optimised \texttt{box\_size} and \texttt{step\_size} inputs for Aegean, PyBDSF, and Selavy,$^a$ using CMP and EXT $\mathcal{D/S}$-image data.}
\centering
\begin{tabular}{@{\;\;}l@{\;\;}c@{\;\;}c@{\;\;}c@{\;\;}c@{\;\;}}
\hline\hline
Image & \multicolumn{1}{l}{Image} &      $\mu$     & \texttt{box\_size}  & \texttt{step\_size} \\
      & \multicolumn{1}{l}{Depth}  & ($\mu$Jy/beam) & ($^{\prime\prime}$) & ($^{\prime\prime}$) \\
\hline
CMP  & $\mathcal{D}$ & 21.81  & 240 & 120 \\
    & $\mathcal{S}$ & 108.2  & 180 &  88 \\
\hline\hline
EXT & $\mathcal{D}$ & 68.01  & 480 & 240 \\
    & $\mathcal{S}$ & 325.3  & 240 & 120 \\
\hline\hline
\multicolumn{4}{l}{\footnotesize$\!\!^a$Selavy only accepts \texttt{box\_size}.}
\end{tabular}
\label{tb:hydra_2x2_typhon_rms_box_stats}
\end{table}

\begin{table*}[hbt!]
\caption{Typhon run statistics for CMP and EXT images, with SF, image depth, SF RMS parameter ($n_{rms}$ [$\sigma$]), SF island parameter ($n_{island}$ [$\sigma$]), source counts (N), residual RMS [$\mu$Jy/beam], and residual MADFM [$\mu$Jy/beam] columns.$^a$}
\centering
\begin{tabular}{lcccrcc@{\;\;\;\;\;\;}ccrcc}
\hline\hline
SF & \multicolumn{1}{l}{Image} & 
   \multicolumn{5}{c}{CMP Sources} & \multicolumn{5}{c}{EXT Sources}\\
 & \multicolumn{1}{l}{Depth} &
   \multicolumn{1}{c}{$n_{rms}$} & \multicolumn{1}{c}{$n_{island}$} & \multicolumn{1}{c}{N} & RMS & MADFM &
   \multicolumn{1}{c}{$n_{rms}$} & \multicolumn{1}{c}{$n_{island}$} & \multicolumn{1}{c}{N} & RMS & MADFM \\ 
\hline
Aegean   & $\mathcal{D}$ & 3.074 & 3.070 & 6\,016 & 20.0 & 19.0 & 3.074 & 3.070 & 4\,112 & 67.0 & 56.0 \\
Caesar   & $\mathcal{D}$ & 3.074 & 3.000 & 4\,084 & 19.0 & 16.0 & 3.206 & 3.000 & 3\,618 & 54.0 & 44.0 \\
ProFound & $\mathcal{D}$ & 2.412 & 2.409 & 4\,997 & 18.0 & 16.0 & 2.412 & 2.409 & 3\,730 & 52.0 & 43.0 \\
PyBDSF   & $\mathcal{D}$ & 2.809 & 2.806 & 5\,991 & 22.0 & 19.0 & 2.809 & 2.806 & 4\,688 & 106 & 54.0 \\
Selavy   & $\mathcal{D}$ & 3.206 & 3.203 & 3\,225 & 45.0 & 20.0 & 3.206 & 3.203 & 2\,544 & 982 & 58.0 \\
\hline\hline                                             
Aegean   & $\mathcal{S}$ & 3.868 & 3.864 &    747 & 110  & 110  & 3.603 & 3.599 & 1\,287 & 321  & 317  \\
Caesar   & $\mathcal{S}$ & 4.000 & 3.000 &    657 & 109  & 106  & 3.603 & 3.000 & 1\,297 & 310  & 295  \\
ProFound & $\mathcal{S}$ & 3.074 & 3.070 &    642 & 109  & 107  & 2.941 & 2.938 & 1\,138 & 311  & 298  \\
PyBDSF   & $\mathcal{S}$ & 3.735 & 3.732 &    598 & 111  & 109  & 3.338 & 3.335 & 1\,312 & 316  & 313  \\
Selavy   & $\mathcal{S}$ & 4.000 & 3.996 &    427 & 114  & 110  & 3.735 & 3.732 &    787 & 623  & 320  \\
\hline\hline
\multicolumn{12}{l}{\footnotesize$\!\!^a$The MADFM estimators are normalised by 0.6744888 \citep{whiting_2012b}.}
\end{tabular}
\label{tb:hydra_2x2_typhon_stats}
\end{table*}

For the CMP source $\mathcal{D}$-image, $\mu\sim22\,\mu$Jy/beam (Table~\ref{tb:hydra_2x2_typhon_rms_box_stats}) which is consistent with the design RMS noise of $20\,\mu$Jy/beam (\textit{re.} \S~\ref{sc:image_data}). For the EXT source $\mathcal{D}$-image, $\mu\sim68\,\mu$Jy/beam which is higher 
due to the inclusion of extended structures with a slightly higher source density, and the fact that the box size has doubled. This is consistent with an average source size increase from $15^{\prime\prime}$ to $30^{\prime\prime}$. We also note that $\mu_{\mathcal{S}}/\mu_{\mathcal{D}}\sim 5$ for all images, which is consistent with the factor of 5 noise increase for the $\mathcal{S}$-images (\S~\ref{sc:homados}). So the RMS box statistics are consistent with what might be expected from the simulated images.

\begin{figure}[htb!]
\begin{center}
\includegraphics[width=\columnwidth]{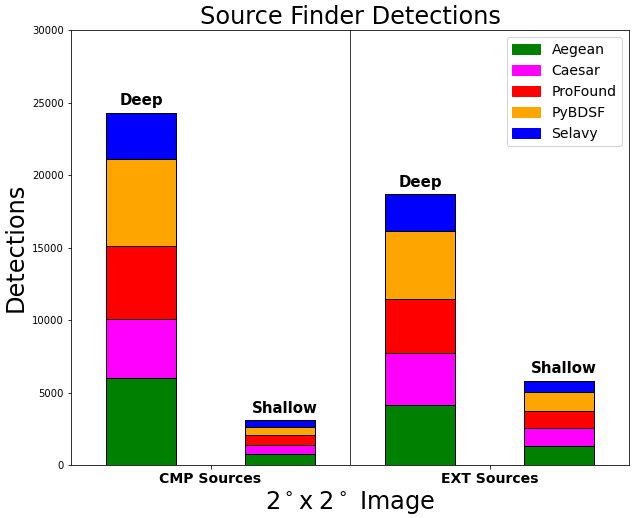}
\caption{SF CMP and EXT $\mathcal{D/S}$-image detection stacked plots (\textit{re.} the N columns of Table~\ref{tb:hydra_2x2_typhon_stats}). \label{fg:source_finder_simulated_detections}}
\end{center}
\end{figure}

In Table~\ref{tb:hydra_2x2_typhon_stats} we notice, in the broadest sense, that the RMS and island parameters are similar for the $\mathcal{D/S}$-images, row-wise. Figure~\ref{fg:source_finder_simulated_detections} shows stacked plots of $\mathcal{D/S}$ source counts for the images. In the CMP and EXT $\mathcal{D}$-images, there is some variability in the source counts, \textit{i.e.} $N\sim4\,650\pm890$, except for Selavy being consistently lower by $N\sim1\,770\pm950$. In the $\mathcal{S}$-image case, the variability is fairly tight, with $N_{\mbox{\tiny CMP}}\sim661\pm54$ and $N_{\mbox{\tiny EXT}}\sim1\,258\pm70$, except for Selavy which is consistently low with $N_{\mbox{\tiny CMP}}=427$ and $N_{\mbox{\tiny EXT}}=787$.

For the CMP $\mathcal{D}$-image, the RMS and MADFM residual statistics are all $\sim 19\,\mu$Jy/beam, with the exception of Selavy having a significantly higher RMS ($\sim45\,\mu$Jy/beam), likely the primary cause for the reduction in numbers of detected sources it reports. For the EXT $\mathcal{D}$-image, the RMS values increase in order from ProFound, Caesar, Aegean, PyBDSF, to Selavy, with the latter three being comparable to the former being at extreme end. This is also reflected in their MAFDMs, although here the values are somewhat comparable. For the $\mathcal{S}$-image case, everything is similar within each image data set, except for Selavy having a higher RMS in the EXT image case. These observed discrepancies for Selavy are likely due to its use of robust statistics in the background estimation (\S~\ref{sc:selavy}). In contrast, the MADFM is similar in both cases for all SFs.

Table~\ref{tb:hydra_2x2_typhon_n_stats} shows ratios of deep-to-injected ($\mathcal{D:J}$) and shallow-to-deep ($\mathcal{S:D}$) source counts: \textit{i.e.}, $\mathcal{D:J}$ indicates the fraction of sources recovered in the simulated images, whereas $\mathcal{S:D}$ indicates the recovery rate assuming the deep detections are \textit{``real''}. Also included are $\mathcal{S:J}$ recovery rates, for comparison.\footnote{As $\mathcal{S:D}=\mathcal{S:J}/\mathcal{D:J}$.\label{fn:as}} The $\mathcal{D:J}$ recovery rates are not expected to reach 100\%, as the simulations includes low S/N sources, and for EXT sources, some lying below S/N$=1$.
The $\mathcal{S:D}$ recovery rate is lower for CMP sources than EXT, which both track consistently with $\mathcal{D:J}$.\footnote{\textit{i.e.},  $(\mathcal{S:J})_{\mbox{\tiny CMP \& EXT}}\sim6.5\pm1.1\%$ implies $\mathcal{S:D}/\mathcal{D:J}$ is roughly constant.$^{\ref{fn:as}}$} This may be due to some confusion in the EXT image (Figure~\ref{fg:extended_source_sim}), given the $\mathcal{S:J}$ recovery rates are similar.

\begin{table}[hbt!]
\caption{Ratios of deep-to-injected ($\mathcal{D:J}$) shallow-to-injected ($\mathcal{S:J}$), and shallow-to-deep ($\mathcal{S:D}$) sources. The $\mathcal{D/S}$ source counts (N) are provided in Table \ref{tb:hydra_2x2_typhon_stats}, and the injected source counts are 9\,075 and 9\,974 for CMP and EXT sources, respectively  (\textit{re.} \S~\ref{sc:image_data}).}
\centering
\begin{tabular}{@{\;}l@{\;\;}c@{\;\;}c@{\;\;}c@{\;\;\;\;}c@{\;\;}c@{\;\;}c@{\;}}
\hline\hline
Source &
   \multicolumn{3}{@{}c@{}}{CMP} & \multicolumn{3}{@{}c@{}}{EXT} \\
Finder & $\mathcal{D:J}$ & $\mathcal{S:J}$ & $\mathcal{S:D}$ 
       & $\mathcal{D:J}$ & $\mathcal{S:J}$ & $\mathcal{S:D}$\\
\hline
Aegean   & 66.3\% & 8.2\% & 12.4\% & 41.2\% & 7.5\% & 31.3\% \\
Caesar   & 45.0\% & 7.2\% & 16.1\% & 36.3\% & 6.6\% & 35.8\% \\
ProFound & 55.1\% & 7.1\% & 12.8\% & 37.4\% & 6.4\% & 30.5\% \\
PyBDSF   & 66.0\% & 6.6\% & 10.0\% & 47.0\% & 6.0\% & 28.0\% \\
Selavy   & 35.5\% & 4.7\% & 13.2\% & 25.5\% & 4.3\% & 30.9\% \\
\hline\hline
\end{tabular}
\label{tb:hydra_2x2_typhon_n_stats}
\end{table}

The consistency of the RMS box, MADFM, $\mathcal{D:J}$, $\mathcal{S:J}$, and $\mathcal{S:D}$ statistics provides a good indication that Hydra's optimisation routines are performing robustly.

\subsubsection{Source Size Distributions}
\begin{figure}[hbt!]
\centering%
\includegraphics[width=\columnwidth]{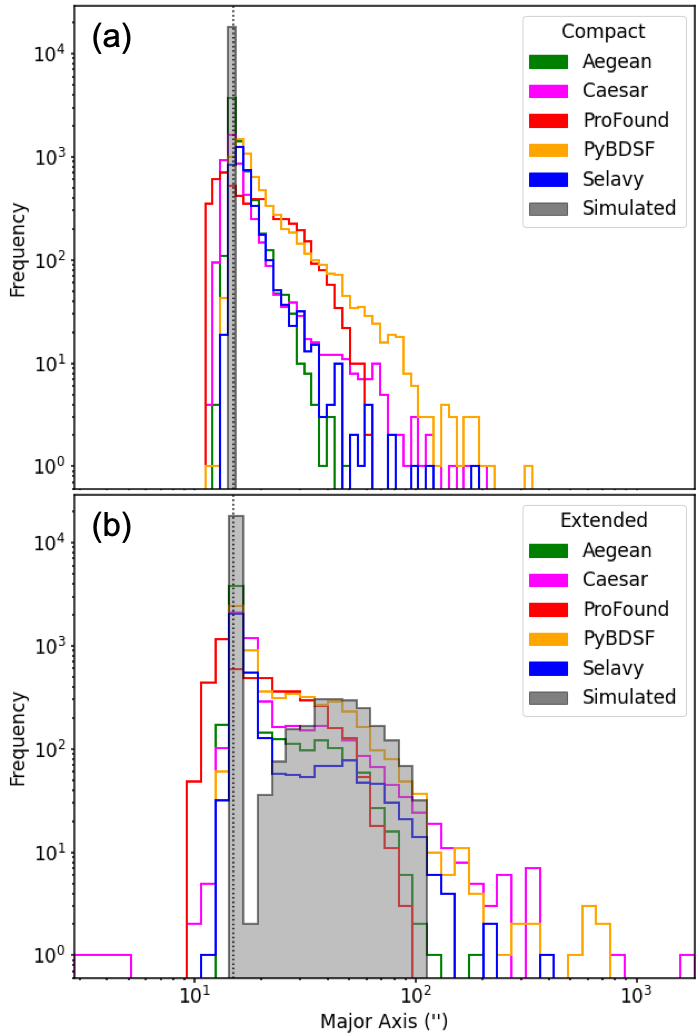}
\caption{Major-axis distributions for (a) CMP and (b) EXT sources (with $\mathcal{D}$ and $\mathcal{S}$ both included). The vertical dashed-line represents the beam size. The distributions of the injected sources are shown in grey. Recall that size estimates between SFs are not necessarily directly comparable as they are estimated using different methods.}
\label{fg:major_distributions}
\end{figure}

Figure~\ref{fg:major_distributions} shows the major-axis distribution for our simulated image data. Both the $\mathcal{D}$ and $\mathcal{S}$ source detections are combined, as the $\mathcal{S}$ provides no contrasting information and its statistics are quite low (Figure~\ref{fg:source_finder_simulated_detections}). Note that the size estimates for different SFs use different methods and are not necessarily directly comparable. Those SFs that fit Gaussians to source components report size as a FWHM, while others (such as ProFound) use different measures, such as a flux-weighted fit (see Appendix~\ref{ap:source_finder_notes} for details).

For the CMP source case, ProFound and PyBDSF tend to overestimate the source sizes. This is likely due to deblending issues or the incorporation of noise spikes. As for EXT sources, PyBDSF tends to most accurately recover the extended source sizes, but, along with Caesar also has the most outliers. These could be attributed to size overestimates due to inclusion of noise spikes or adjacent sources in the fitted sizes. All other SFs tend to marginally underestimate the EXT source sizes. Components smaller than $15^{\prime\prime}$ are attributed primarily to noise spikes, but also occasionally to underestimating the source sizes.

\subsection{Completeness and Reliability}
\label{sc:c_and_r}
In this section, we discuss deep ($\mathcal{D}$) and shallow ($\mathcal{S}$) completeness ($\mathcal{C}$) and reliability ($\mathcal{R}$) (see \S~\ref{sc:oh_goody}). We then provide justification for using deep-shallow ($\mathcal{DS}$) completeness ($\mathcal{C_{DS}}$) and reliability ($\mathcal{R_{DS}}$) metrics for real images, and qualifications on their use, through goodness of completeness ($\mathcal{\tilde{C}_{DS}}$) and goodness of reliability ($\mathcal{\tilde{R}_{DS}}$) results.

\subsubsection{Simulated Sources}
\label{sc:cd_and_rs}
\begin{figure*}[hbt!]
\centering%
\includegraphics[width=2\columnwidth]{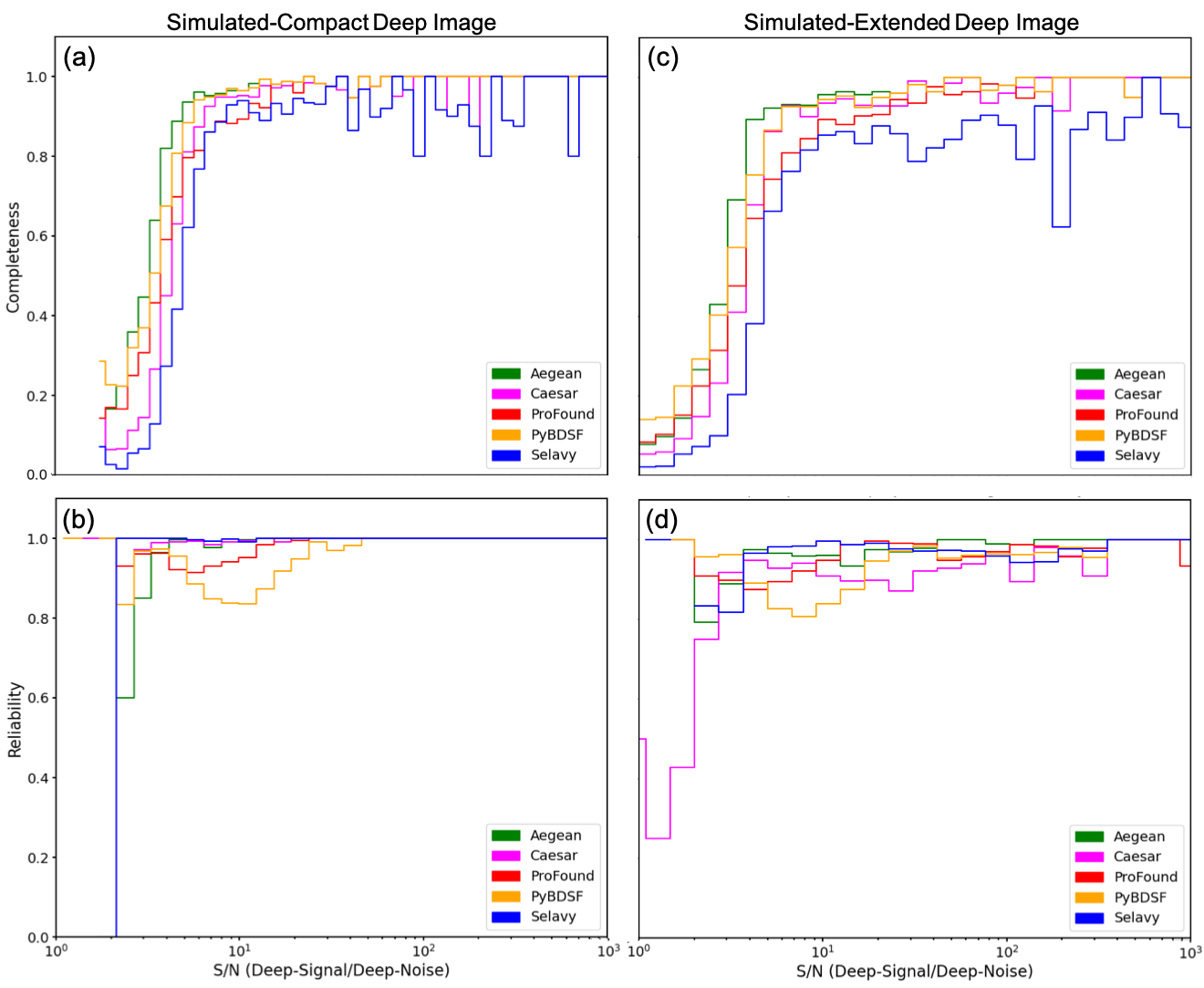}
\caption{Simulated $\mathcal{D}$-image compact (left) and extended (right) source $\mathcal{C_{D}}$ (top) and $\mathcal{R_{D}}$ (bottom) \textit{vs.} S/N ($\mathcal{D}$-signal/Deep-noise) plots. The $\mathcal{D}$-noise is computed using \textsc{bane}.}
\label{fg:simulated_deep_cr}
\end{figure*}

\begin{figure*}[hbt!]
\centering%
\includegraphics[width=2\columnwidth]{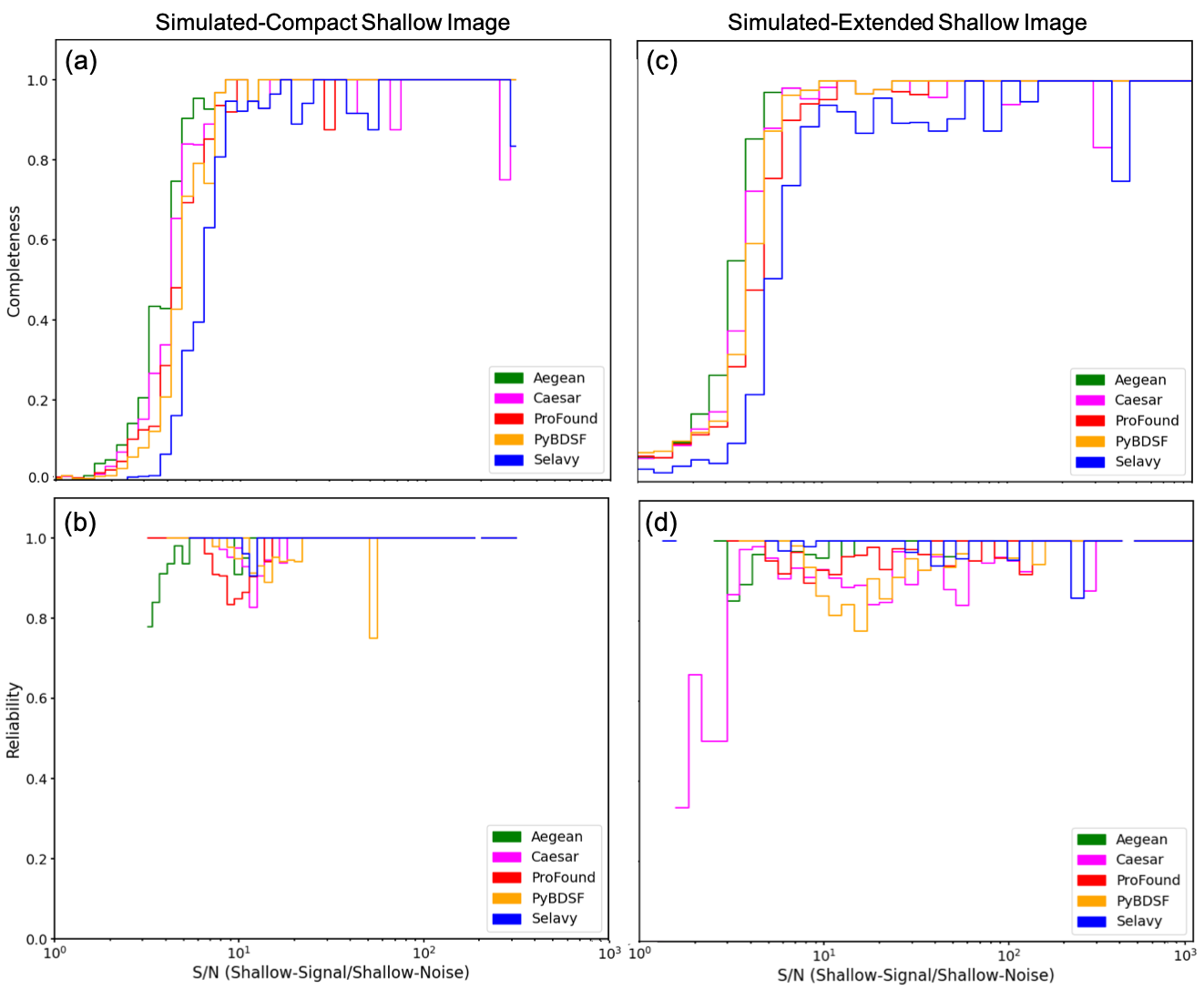}
\caption{Simulated $\mathcal{S}$-image compact (left) and extended (right) source $\mathcal{C_{S}}$ (top) and $\mathcal{R_{S}}$ (bottom) \textit{vs.} S/N ($\mathcal{D}$-signal/$\mathcal{S}$-noise) plots. The $\mathcal{S}$-noise is computed using \textsc{bane}.}
\label{fg:simulated_shallow_cr}
\end{figure*}

Figure~\ref{fg:simulated_deep_cr} shows $\mathcal{C_{D}}$ \textit{vs.} S/N (top) and $\mathcal{R_{D}}$ \textit{vs.} S/N (bottom) for CMP (left) and EXT (right) $\mathcal{D}$-images, where the signal-to-noise (S/N) is the ratio of the $\mathcal{D}$-signal to $\mathcal{D}$-noise. Figure~\ref{fg:simulated_shallow_cr} shows the corresponding results for $\mathcal{S}$-images, where S/N is the ratio of the $\mathcal{S}$-signal to $\mathcal{S}$-noise.

For CMP sources all SFs show a similar behaviour in general. The completeness metric starts to decline slowly from 100\% at S/N$\lesssim 30$, and drops rapidly toward zero below S/N$\sim 5$. False detections are typically limited to about 10\% of the sample down to S/N$\sim 5$, but can appear in some SFs as high as S/N$\sim 30$. In general terms, Aegean provides the highest level of completeness at any given S/N, with a well-behaved decline in reliability below S/N$\sim 5-6$. At the other end of the scale, Selavy has the lowest level of completeness at any S/N, but the best reliability (fewest false detections).

All SFs seem to miss some bright sources, with Selavy standing out as the poorest in this regard. This is likely due in part to the handling of overlapping sources. Both PyBDSF and ProFound report the largest numbers of false sources (seen in $\mathcal{R_D}$ and $\mathcal{R_S}$) at high S/N. For PyBDSF this is a consequence of overestimating source sizes (Figure~\ref{fg:major_distributions}), especially in the presence of closely neighbouring sources, or nearby noise spikes, by quite significant amounts in some cases (see Paper~II). For ProFound this arises due to the blending of neighbouring sources (see Paper~II). 

For EXT sources there is generally poorer performance overall compared to those for the CMP sources, most clearly seen in the $\mathcal{D}$-image results (Figure~\ref{fg:simulated_deep_cr}). There are also several artifacts appearing at unphysically low S/N ($\mbox{S/N}<1$) arising from spurious faint detections. Even at reasonable S/N ($10<\mbox{S/N}<100$) there is measurable incompleteness, and reliability that dips as low as 80\% for some SFs. Here Aegean appears to perform the best, with Selavy showing much poorer performance.

\subsubsection{Metrics for Real Sources}
\label{sc:cds_and_rds}
\begin{figure*}[hbt!]
\centering%
\begin{tabular}{c}
\includegraphics[width=0.95\textwidth]{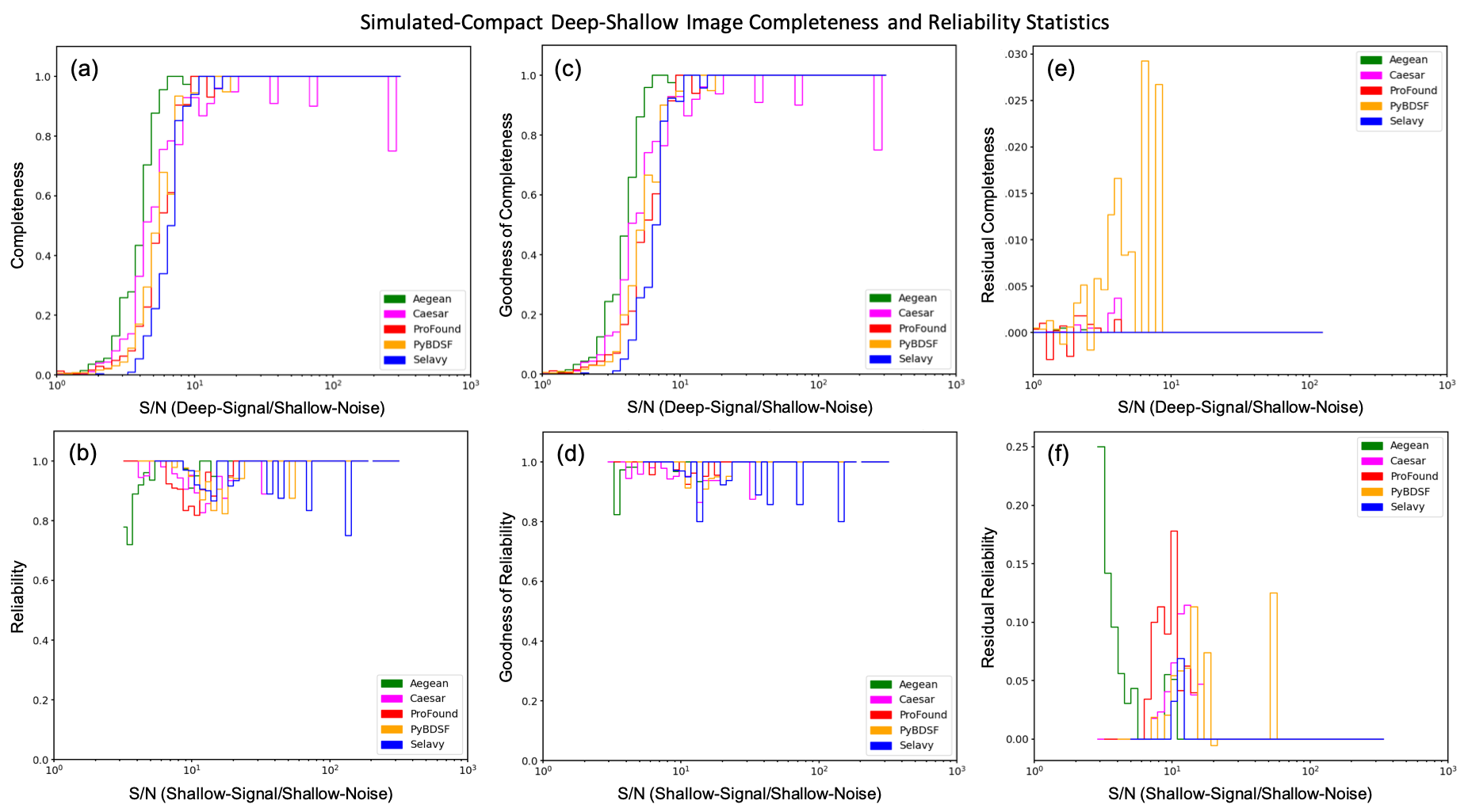}\\
\includegraphics[width=0.95\textwidth]{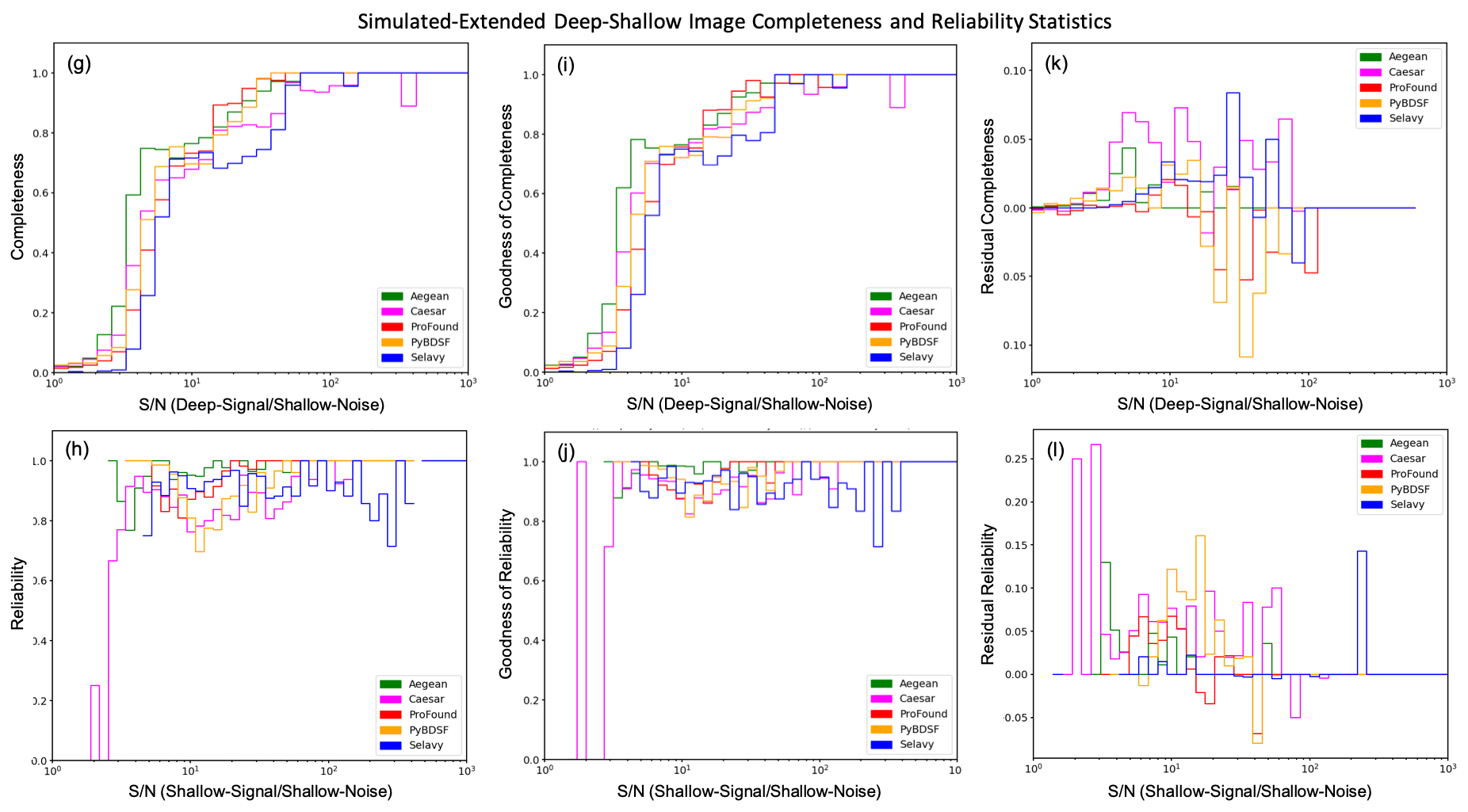}
\end{tabular}
\caption{$\mathcal{C_{DS}}$ (a and g), $\mathcal{R_{DS}}$  (b and h), $\mathcal{\tilde{C}_{DS}}$ (c and i), $\mathcal{\tilde{R}_{DS}}$ (d and j), $\delta\mathcal{C_{DS}}$ (e and k), and $\delta\mathcal{R_{DS}}$ (f and l) \textit{vs.} S/N for CMP (top set) and simulated-extend (bottom set) sources. The S/N are expressed as $\mathcal{D}$-signal/$\mathcal{S}$-noise and $\mathcal{S}$-signal/$\mathcal{S}$-noise for completeness and reliability, respectively, where the $\mathcal{S}$-noise is computed using \textsc{bane}.}
\label{fg:dj_ds}
\end{figure*}

Figure~\ref{fg:dj_ds} (a) and (b) show $\mathcal{C_{DS}}$ \textit{vs.} S/N and $\mathcal{R_{DS}}$ \textit{vs.} S/N, respectively, for CMP sources. As we know what the true sources are, we can explore the validity of $\mathcal{C_{DS}}$ and $\mathcal{R_{DS}}$ by removing any false detections in the $\mathcal{D}$ and $\mathcal{S}$-images, \textit{i.e.}, we can compute $\mathcal{\tilde{C}_{DS}}$ (Equation~\ref{eq:godness_of_completeness}) and $\mathcal{\tilde{R}_{DS}}$ (Equation~\ref{eq:godness_of_reliability}). The results are shown in Figure~\ref{fg:dj_ds} (c) and (d), respectively.

Comparing $\mathcal{\tilde{C}_{DS}}$ and $\mathcal{\tilde{R}_{DS}}$ with $\mathcal{C}_{\mathcal{DS}}$ and $\mathcal{R}_{\mathcal{DS}}$, respectively, $\tilde{\mathcal{C}}_{\mathcal{DS}}$ is largely unchanged, while $\tilde{\mathcal{R}}_{\mathcal{DS}}$ does not show the dip in $\mathcal{R}_{\mathcal{DS}}$ around S/N $\sim$ 10 seen for most SFs. It also excludes the decline seen by Aegean at low S/N. This suggests that the apparently poorer estimated reliability in $\mathcal{R}_{\mathcal{DS}}$ arises from the existence of spurious sources detected in the $\mathcal{D}$-image that are (not surprisingly) missed in the $\mathcal{S}$-image. 

To quantify these results we define residuals between our unconstrained ($\mathcal{C}_{\mathcal{DS}}$ and $\mathcal{R}_{\mathcal{DS}}$) and constrained ($\mathcal{\tilde{C}_{DS}}$ and $\mathcal{\tilde{R}_{DS}}$) metrics: \textit{i.e.}, the residual completeness,
\begin{equation}
    \delta\mathcal{C_{DS}}=\tilde{\mathcal{C}}_{\mathcal{DS}}-\mathcal{C}_{\mathcal{DS}}\,, \label{eq:residual_completeness}
\end{equation} 
and residual reliability,
\begin{equation}
    \delta\mathcal{R_{DS}}=\tilde{\mathcal{R}}_{\mathcal{DS}}-\mathcal{R}_{\mathcal{DS}}\,. \label{eq:residual_reliability}
\end{equation} 
These quantities are shown in Figure~\ref{fg:dj_ds} (e) and (f), respectively. In general, they are expected to be positive, as there should be an excess of false detections in $\mathcal{C_{DS}}$ and $\mathcal{R_{DS}}$ compared to $\mathcal{\tilde{C}_{DS}}$ and $\mathcal{\tilde{R}_{DS}}$, by construction. Negative values may appear when real input sources are detected but poorly fit, leading to inconsistent flux densities between the $\mathcal{D}$ and $\mathcal{S}$-images.  
The figure shows $\delta\mathcal{C_{DS}}$ predominantly highlighting a small number of spurious PyBDSF detections, and $\delta\mathcal{R_{DS}}$ emphasising the distribution in S/N of the false detections.

Recognising these limitations, while also noting that for several SFs $\delta\mathcal{C_{DS}}$ and $\delta\mathcal{R_{DS}}$ are small, the approach of estimating completeness and reliability for a given finder based on real images is not unreasonable. Clearly it is not as robust as doing so using known injected sources as a reference, but it may be a useful addition to analyses comparing finder performance on real data containing imaging artifacts and other hard to simulate systematics. Similar conclusions can be drawn for the EXT source case (Figure~\ref{fg:dj_ds} (g) through (l)).

Finally, we note that $\mathcal{C_{DS}}$ and $\mathcal{R_{DS}}$ in Figure~\ref{fg:dj_ds} (a) and (b), respectively, for CMP sources fare much better than their EXT counterparts, (g) and (h), respectively. The metrics in the latter case perform even more poorly than their  $\mathcal{C_{D}}$, $\mathcal{R_{D}}$, $\mathcal{C_{S}}$, and $\mathcal{R_{S}}$ counterparts (Figures~\ref{fg:simulated_deep_cr} and~\ref{fg:simulated_shallow_cr}). This could perhaps be due to confusion, given the source density (Figure~\ref{fg:extended_source_sim}) and high MADFM statistics (Table~\ref{tb:hydra_2x2_typhon_stats}).

\subsection{Summary of validation tests}
The Hydra software was tested using the Aegean, Caesar, ProFound, PyBDSF, and Selavy SFs, comparing them by first minimising the FDR based on a 90\% PRD cutoff, through Typhon.
This process was done for $2^\circ\times2^\circ$ CMP and EXT $\mathcal{D/S}$-images. The RMS box, MADFM, and source detection statistics were also shown to be consistent with the simulated data, thus validating Hydra's performance. The source size distributions also provided an indication of its performance.

The simulated data was used to develop $\mathcal{C_{DS}}$ and $\mathcal{R_{DS}}$ metrics for deep/shallow image pairs, treating $\mathcal{D}$ detections as true sources. This was done by examining these statistics with the erroneous detections filtered out, using knowledge of the underlying injected sources, leaving goodness of completeness, $\mathcal{\tilde{C}_{DS}}$, and goodness of reliability, $\mathcal{\tilde{R}_{DS}}$, metrics. Contrasting  $\mathcal{C_{DS}}$ and $\mathcal{R_{DS}}$ with $\mathcal{C_D}$ (or $\mathcal{C_S}$) and $\mathcal{R_D}$ (or $\mathcal{R_S}$), respectively, a notable degradation in the former was observed for EXT images, most likely due to $\mathcal{DS}$-confusion. That being said, in general, the form of $\mathcal{C_{DS}}$ and $\mathcal{R_{DS}}$ remains relatively unchanged, although the $\mathcal{S}$ recovery rate, $\mathcal{S:D}$, is significantly reduced (Table~\ref{tb:hydra_2x2_typhon_n_stats}). This suggests that these metrics are good for studying SF performance in real images, given the ability to quantify incompleteness ($1-\mathcal{C}$) and FDR ($1-\mathcal{R}$).

In passing, some observations of SF performance were also made, and are explored in more detail in Paper~II. The source detection numbers were comparable between all SFs except for Selavy, which was consistently low (Table~\ref{tb:hydra_2x2_typhon_n_stats}). Some variability in the residual RMS estimated for the $\mathcal{D}$-mages was observed, with Selavy having unusually high values. For CMP sources the values were comparable except for Selavy, whereas for the EXT case there was significant variation except for Caesar and Profound. As for the $\mathcal{S}$-images the values were all consistent, except for Selavy being significantly high in the EXT source case. The MAFDM statistics were consistent in both cases for all SFs.

For CMP and EXT sources, Aegean had the best $\mathcal{C_D}$ (and $\mathcal{C_S}$) statistics followed by PyBDSF, ProFound, Caesar, and Selavy. Selavy, followed to a lesser degree by Caesar, tends to miss bright sources, more so than the other SFs. For $\mathcal{R_D}$ (and $\mathcal{R_S}$) they also reported the largest number of false sources at high S/N. The quality of the $\mathcal{C}$ and $\mathcal{R}$ statistics is poorer for EXT than CMP sources which is mainly attributed to confusion (see Figure~\ref{fg:extended_source_sim}).

\section{Conclusions}
\label{sc:summary}
Radio astronomy has dramatically progressed in the lead-up to the SKA era \citep[see, \textit{e.g.}, Figure~1 of][]{norris_2021}. The past decades have seen development of technologies and facilities that improve the survey speed, survey depth, and a rapid growth in results from SKA precursors. Ongoing and planned surveys such as VLASS \citep{lacy_2020,gordon_2021} with 82\% sky-coverage, and EMU \citep{norris_2011, norris_2021} with 75\% sky-coverage, expect to produce catalogues with source numbers into the millions and tens of millions.

Increasing survey sizes drive a need for SF software with highly robust and well-characterised completeness and reliability statistics. This need has driven source-finding challenges \citep[\textit{e.g.,}][]{hopkins_2015,bonaldi_2021} for comparing the various tools and technologies. Some optical SFs have also been applied to radio images, such as SExtractor \citep{bertin_1996} and ProFound \citep{hale_2019}. Caesar was introduced for handling compact sources jointly with diffuse emission \citep{umana_2015}, through the reprocessing of its residual image \citep{riggi_2016}. These are natural extensions of traditional thinking on radio source extraction. Qualitatively different approaches are also being developed, including the application of machine learning to this field \cite[\textit{e.g.,}][]{bonaldi_2021,magro_2022}. In the SKA era, it may be that source detection and cataloguing will need to be done on the fly due to the data volume \citep{bonaldi_2021}. A Hydra-like tool may have substantial value in that context, encapsulating the strengths of multiple SFs run in parallel.

The optimum comparison between SFs requires the expertise of the originators to fine tune their performance for a given set of reference images, as pursued in such data challenges. For this reason it is necessary to fairly compare large numbers of SFs on an even footing. Hydra was developed to encapsulate this expertise, in a modular fashion, using Docker containers. Hydra is extensible and the user does not have to be an expert at every SF to use it. Hydra focuses specifically on optimising the RMS threshold and island growth parameters, common to the traditional class of SFs, through the percentage real detections metric, PRD (Equation~\ref{eq:prd}). This two-parameter optimisation technique is adopted following \cite{hale_2019}, who did the optimisation by hand, in a study comparing Aegean, ProFound, and PyBDSF. Hydra also includes an optional background-estimation optimisation step to identify the RMS box and step size parameters, through the mean noise metric, $\mu$ (Equation~\ref{eq:rms_box_pars}). It is clearly possible to improve Hydra in order to handle more SF-specific parameters.

Hydra provides deep ($\mathcal{D}$, \textit{i.e} the input image) and shallow ($\mathcal{S}$, \textit{i.e.}, the $D$-image with $5\sigma$ noise added) catalogues for each SF, which are linked through a cluster table of overlapping components, or clumps (Figure~\ref{fg:clustering_infofographic}). The $\mathcal{S}$-image creation and analysis is motivated by a desire to assess a given finder's performance against itself, in the absence of simulated data, treating the $\mathcal{D}$-image version of the SF's catalogue as the ground truth. Each clump has an ID which can be used to locate the associated $\mathcal{D/S}$-image and residual cutouts (with and without component annotations). $\mathcal{D/S}$ region files for the full images are also available. Hydra also merges catalogues of known sources (simulated in our case), providing corresponding metrics such as completeness ($\mathcal{C}$) and reliability ($\mathcal{R}$). Hydra also comes with an HTML viewer that allows the user to explore the various data products (Figure~\ref{fg:hydra_viewer}).

This paper is part one of a two part series, in which we have introduced the Hydra software, and validated its optimisation algorithms, using simulated-compact (CMP) and simulated-extended (EXT) image data. In addition to the traditional $\mathcal{D}$-image metrics, such as completeness ($\mathcal{C_D}$) and reliability ($\mathcal{R_D}$), Hydra introduces a whole new set of metrics, such as $\mathcal{S}$-image completeness ($\mathcal{C_S}$) and reliability ($\mathcal{R_S}$), and $\mathcal{DS}$-image completeness ($\mathcal{C_{DS}}$) and reliability ($\mathcal{R_{DS}}$), respectively. In this paper we also validated the $\mathcal{C_{DS}}$ and $\mathcal{R_{DS}}$ metrics for use with real images, by using our simulated data where the true sources are known. It was found that $\mathcal{C_{DS}}$ and $\mathcal{R_{DS}}$ are useful for characterising SF performance, provided one keeps in mind the $\mathcal{D}$ detections are incomplete with a slight degradation in $\mathcal{R_{DS}}$ for low S/N. In Paper~II we evaluate the performance of the SFs using our simulated images along with real data.

\section{Acknowledgements}
M. M. Boyce, S. A. Baum, Y. A. Gordon, D. Leahy, C. O'Dea, and A. N. Vantyghem acknowledge partial support from the NSERC of Canada. S. Riggi acknowledges INAF for financial support through the PRIN TEC programme ``CIRASA.'' Partial support for L. Rudnick came from U.S. National Science Foundation Grant AST17-14205 to the University of Minnesota. M. I. Ramsay acknowledges support from NSERC and the University of Manitoba Faculty of Science Undergraduate Research Award (USRA). C. L. Hale acknowledges support from the Leverhulme Trust through an Early Career Research Fellowship. Y. A. Gordon is supported by US National Science Foundation grant 2009441. H. Andernach benefited from grant CIIC 138/2020 of Universidad de Guanajuato, Mexico. D. Leahy acknowledges support from NSERC 10020476. M. J. Micha{\l}owski acknowledges the support of the National Science Centre, Poland through the SONATA BIS grant 2018/30/E/ST9/00208. S. Safi-Harb acknowledges support from NSERC through the Discovery Grants and the Canada Research Chairs programs, and from the Canadian Space Agency. M. Vaccari acknowledges financial support from the Inter-University Institute for Data Intensive Astronomy (IDIA), a partnership of the University of Cape Town, the University of Pretoria, the University of the Western Cape and the South African Radio Astronomy Observatory, and from the South African Department of Science and Innovation's National Research Foundation under the ISARP RADIOSKY2020 Joint Research Scheme (DSI-NRF Grant Number 113121) and the CSUR HIPPO Project (DSI-NRF Grant Number 121291). E. L. Alexander gratefully acknowledges support from the UK Alan Turing Institute under grant reference EP/V030302/1 and from the UK Science \& Technology Facilities Council (STFC) under grant reference ST/P000649/1. A. S. G. Robotham  acknowledges support via the Australian Research Council Future Fellowship Scheme (FT200100375). H. Tang acknowledges the support from the Shuimu Tsinghua Scholar Program of Tsinghua University.

Hydra is written primarily in Python, with some elements of Cython \citep{behnel_2011} and R, along with their standard libraries. Hydra uses alphashape \citep{bellock_2021}, APLpy \citep{robitaille_2012}, Astropy \citep{astropy_2013, astropy_2018}, Matplotlib \citep{hunter_2007}, NumPy \citep{harris_2020}, and pandas \citep{mckinney_2010,reback_2020} Python libraries commonly used in astronomy. Hydra utilizes click, gzip, Jinja, tarfile, and YAML Python libraries as part its overall architectural infrastructure. Hydra encapsulates Aegean \citep{hancock_2012,hancock_2018}, Caesar \citep{riggi_2016,riggi_2019}, ProFound \citep{robotham_2018,hale_2019}, PyBDSF \citep{mohan_2015}, and Selavy \citep{whiting_2012} source finder software using Docker. The technical diagrams in this paper we created using macOS Preview and Microsoft PowerPoint. We acknowledge the authors of the aforementioned software applications, languages, and libraries.

M. M. Boyce would like to thank T. Galvin for discussions regarding clustering techniques (\textit{re.} \S~\ref{sc:hydra_sw}) and extending these concepts to potential computational geometry applications \citep[\textit{re.},][]{edelsbrunner_2013} in multiwavelength astronomy.

CIRADA is funded by a grant from the Canada Foundation for Innovation 2017 Innovation Fund (Project 35999) and by the Provinces of Ontario, British Columbia, Alberta, Manitoba and Quebec, in collaboration with the National Research Council of Canada, the US National Radio Astronomy Observatory and Australia’s Commonwealth Scientific and Industrial Research Organisation.

The National Radio Astronomy Observatory is a facility of the National Science Foundation operated under cooperative agreement by Associated Universities, Inc.

ASKAP is part of the ATNF which is managed by the CSIRO. Operation of ASKAP is funded by the Australian Government with support from the National Collaborative Research Infrastructure Strategy. ASKAP uses the resources of the Pawsey Supercomputing Centre. Establishment of ASKAP, the Murchison Radio-astronomy Observatory and the Pawsey Supercomputing Centre are initiatives of the Australian Government, with support from the Government of Western Australia and the Science and Industry Endowment Fund. We acknowledge the Wajarri Yamatji people as the traditional owners of the Observatory site.

\begin{appendix}
\section{Cerberus Code Template Notes}
\label{ap:cerberus_detail}
In this appendix we provide more architectural details regarding Cerberus template rules discussed in \S~\ref{sc:cerberus}. The intent here is to provide enough detail to give an overall sense of Hydra's extensible nature. Further details can be found in the Hydra user manual. 

\begin{figure*}[htb!]
\begin{center}
\includegraphics[width=\textwidth]{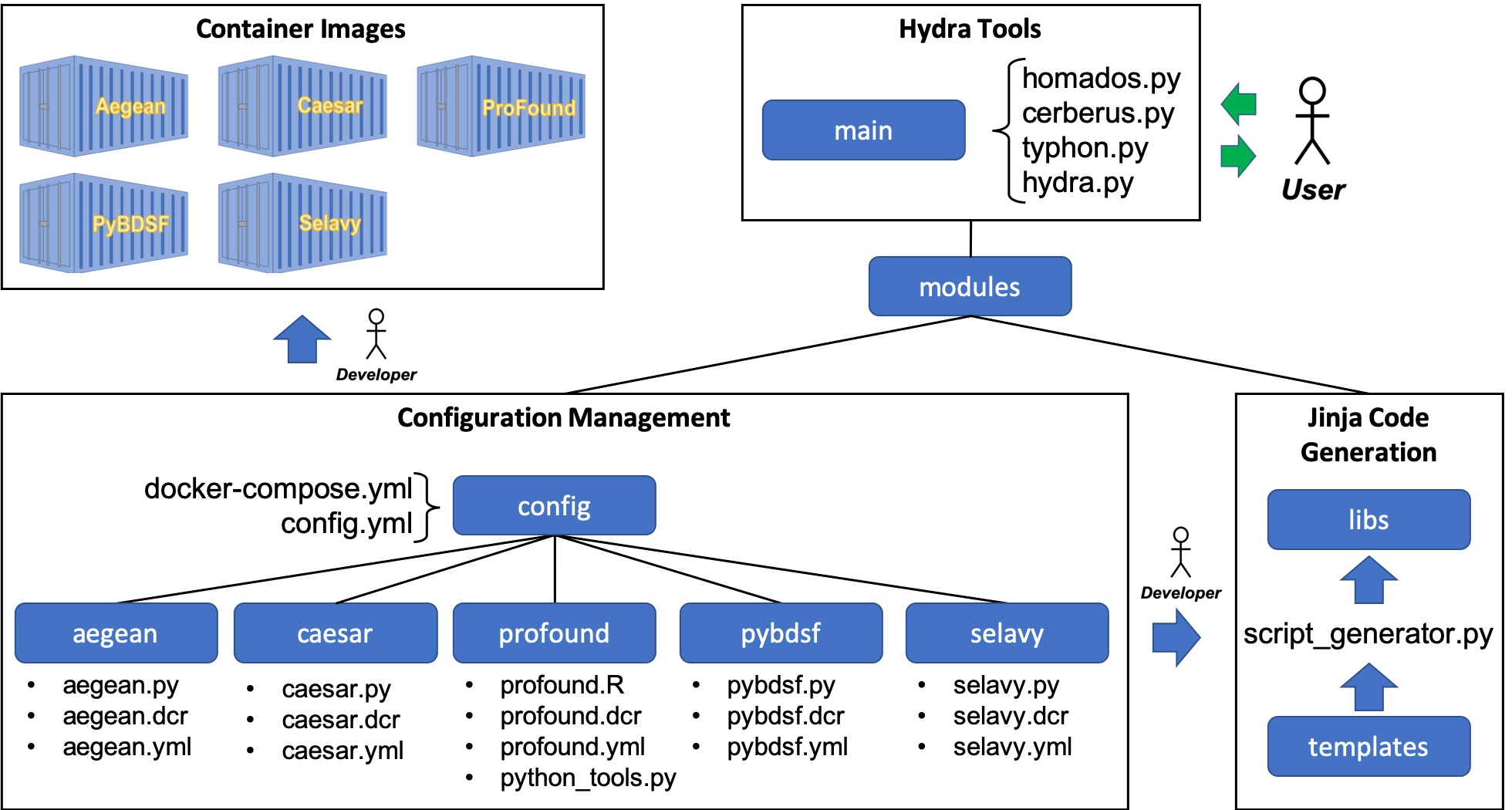}
\caption{Detailed breakdown of the Cerberus code generation workflow given in Figure~\ref{fg:Cerberus_Code_Generation}. Figure~\ref{fg:aegean_containerisation} shows a detailed example of an Aegean container image build, utilising \texttt{aegean.dcr}, \texttt{aegean.py}, and \texttt{docker-compose.yml}. Figure~\ref{fg:aegean_code_generation} shows an example of updating \texttt{cerberus.py} to include Aegean, through code generation, utilising \texttt{aegean.yml} and \texttt{config.yml}. All of the information in Configuration Management is accessible to all of the tools within the Hydra software suite.}
 \label{fg:cerberus_detail}
\end{center}
\end{figure*}

Figure~\ref{fg:cerberus_detail} shows a more detailed view of the Cerberus code generation workflow (\textit{c.f.,} Figure~\ref{fg:Cerberus_Code_Generation}). The term \textit{``template''} refers to the overall directory hierarchy, configuration files, and naming conventions. At the lowest level, within the \texttt{config} directory, are subdirectories for each of the SFs, containing \texttt{*.py} and/or \texttt{*.R} script files, \texttt{*.dcr} Dockerfiles, and \texttt{*.yml} YAML files. The \texttt{docker-compose.yml} and \texttt{config.yml} files in the main \texttt{config} directory provide the glue for building containers and code generation, respectively.

\subsection{Containerisation}
\label{ap:containerisation}
The general recipe for containerising SFs is as follows.
{\small\begin{itemize}
    \item Create a Docker build file containing the following:
    \begin{itemize}
        \item Base operating system environment
        \item SF environment with tools
        \item SF wrapper script with command-line arguments:
        \begin{itemize}
            \item Input image path
            \item Processing directory path
            \item Output directory path
            \item Image filename to process
            \item RMS-Parameter with default setting
            \item Island-Parameter with default setting
            \item \textit{RMS box parameters (optional)}
            \item FITS catalogue file output flag (default, CSV)
            \item Residual image flag
            \item Dump flag
            \item Help flag
        \end{itemize}
        \item Internal directory structure:
        \begin{itemize}
            \item Script home directory\footnote{Used for software development and testing.}
            \begin{itemize}
                \item Input subdirectory
                \item Processing subdirectory
                \item Results subdirectory
            \end{itemize}
        \end{itemize}
        \item A container \texttt{ENTRYPOINT}
    \end{itemize}
    \item Update the \texttt{docker-compose.yml} configuration file with the container build instructions
    \item Build the container image
\end{itemize}}

\begin{figure*}[htb!]
\begin{center}
\includegraphics[width=\textwidth]{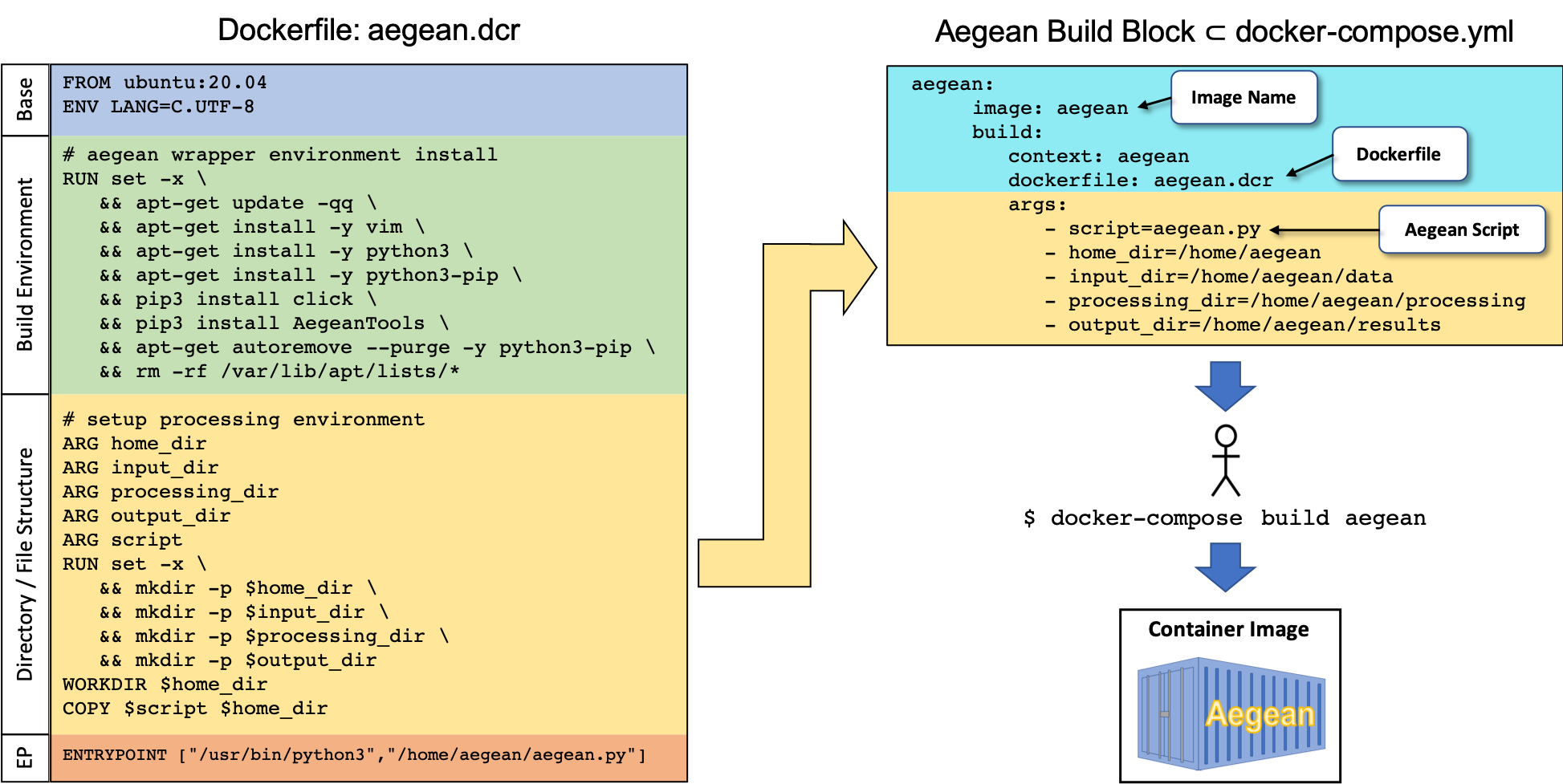}
\caption{Example of Aegean containerisation. The Dockerfile, \texttt{aegean.dcr}, has four main parts, (1) a base Ubuntu 20.04 operating system, (2) an \texttt{AegeanTools} toolbox build environment, (3) a \texttt{home\_dir}, \texttt{input\_dir}, \texttt{processing\_dir}, and \texttt{output\_dir} directory structure along with a local \texttt{script}, \texttt{aegean.py}, and (4) an \texttt{ENTRYPOINT} (EP) through which \texttt{aegean.py} can be externally accessed by \texttt{cerberus.py}. The \texttt{docker-compose.yml} configuration file contains an Aegean build block (\texttt{aegean:}), which has two main parts, (1) a part defining the image name (\texttt{aegean}) along with a pointer the Dockerfile (\texttt{aegean/aegean.dcr}), and (2) a part containing the directory file structure to be built along with a pointer to \texttt{aegean.py}. The container image is built, using this information, with the \texttt{docker-compose} command.}
 \label{fg:aegean_containerisation}
\end{center}
\end{figure*}

\noindent The input and output directories serve as external mount points, used by the \texttt{cerberus.py} wrapper script: \textit{i.e.}, the input directory contains the input image, the processing directory contains the SF wrapper script scratch files, and the results directory contains output catalogues, region files, \textit{etc.}\footnote{The dump flag (\texttt{-\,-dump}) copies the contents of the internal processing directory to the external results directory, which can be used for debugging purposes.} The container \texttt{ENTRYPOINT} allows \texttt{cerberus.py} external access to the internal script. Aegean is perhaps one of the simplest SFs to containerise, with details shown in Figure~\ref{fg:aegean_containerisation}. 

As can be seen, Aegean comes as part of an \texttt{AegeanTools} toolbox within an Ubuntu 20.04 operating system. The Dockerfile, container directory structure, and local \texttt{aegean.py} container wrapper script are defined in the \texttt{docker-compose.yml} configuration file. Everything related to Aegean containers, directories, scripts, \textit{etc.}, are all prefixed with \texttt{aegean}. Also, \texttt{aegean.py} can be accessed internally within the container, \textit{e.g.},

{\footnotesize\begin{verbatim}
/home/aegean# python3 aegean.py --help
Usage: aegean.py [OPTIONS] \
       INPUT_DIR \
       PROCESSING_DIR \
       OUTPUT_DIR FITS_IMAGE_FILE
  Aegean image processing tool.
  inputs:
        INPUT_DIR: location of image_filename.fits
        PROCESSING_DIR: location of scratch directory
        OUTPUT_DIR: location to place results
        FITS_IMAGE_FILE: image_filename.fits
  outputs:
        OUTPUT_DIR/image_filename.aegean.csv
        OUTPUT_DIR/image_filename.aegean.reg
Options:
  --seedclip FLOAT    Island seeding parameter.
  --floodclip FLOAT   Island growing parameter.
  --box-size INTEGER  RMS Box Size (requires: step-size).
  --step-size INTEGER RMS Step Size (requires: box-size).
  --fits              Output FITS catalogue.
  --residual          Output residual and module files.
  --dump              Dump out all processing files.
  --help              Show this message and exit.
/home/aegean#
\end{verbatim}}
\noindent or externally outside of the container, \textit{i.e.},
{\footnotesize\begin{verbatim}
$ docker run --rm -t aegean --help
\end{verbatim}}
\noindent the latter being used by \texttt{cerberus.py} (\S~\ref{sc:cerberus}). After implementing the above template rules a new container image can be built using the \texttt{docker-compose} command within the \texttt{config} directory.

\subsection{Code Generation}
\label{ap:code_generation}
\begin{figure}[htb!]
\begin{center}
\includegraphics[width=\columnwidth]{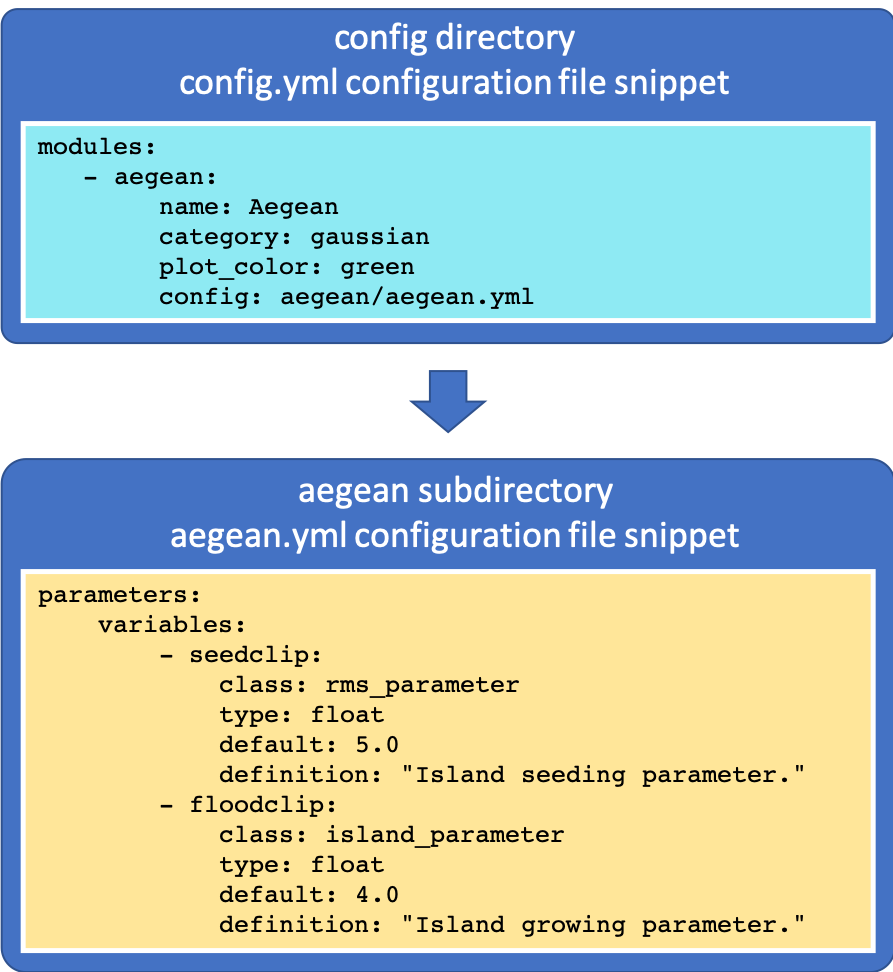}
\caption{Example of the addition of an Aegean module for \texttt{cerberus.py} through code generation. In short, the developer creates an \texttt{aegean.yml} metadata file and links it to the master configuration file, \texttt{config.yml}. Then by running the \texttt{script\_generator.py} script, in the \texttt{libs} directory, the module is installed.}
 \label{fg:aegean_code_generation}
\end{center}
\end{figure}

For code generation a metadata file needs to be created (\textit{e.g.}, \texttt{aegean.yml}), and then linked to the master configuration file, \texttt{config.yml}. This information is then used for code generation through the Jinja template engine. Figure~\ref{fg:aegean_code_generation} shows an example workflow for creating the Aegean module.

All scripts within the Hydra software suite have access to the Configuration Management system in order to perform operations in a generic fashion. For example, \texttt{cerberus.py} utilises the docker-compose configuration file for linking calls to the source-finder container images, \texttt{typhon.py} utilises the metadata files for parameters and constraints used for source-finder optimisation, \texttt{hydra.py} utilises the metadata files for catalogue processing, and so on.

\section{Source Finder Implementation Notes}
\label{ap:source_finder_notes}
In this section we briefly overview the SFs currently supported by Hydra and their relevant settings. 

It should be noted that Aegean, Caesar, and Selavy have various multiprocessor mode implementations wherein large images are split into manageable chunks and processed in parallel in order to reduce the overall processing time \citep[see][respectively for details]{hancock_2018,riggi_2019,whiting_2012}. For the current implementation of Hydra we have chosen to leave these modes disabled. These modes provide various methods for dealing with background noise computations and source detection, which become problematic near edges of sub-images, and consequently have tendencies to bias the statistics especially when comparing against non-multiprocessor SFs, such as ProFound and PyBDSF.

\subsection{Aegean}
The \texttt{AegeanTools} toolbox contains two main items of relevance to this discussion, a background and RMS noise computation script, \textsc{bane}, and a source finding script, \texttt{Aegean} \citep{hancock_2018}. \textsc{bane} uses a sliding box-car method, with grid-based box and step size parameters, wherein RMS noise estimates are calculated using sigma-clipping. Aegean itself uses a faster, but less accurate, ``zones'' algorithm. Aegean can also use the output from \textsc{bane}, which is the implementation adopted here.

The Aegean SF uses a flood-fill algorithm to identify islands above a detection threshold. It then implements a deblending process to determine the number of local maxima through the discrete 2D Laplacian (\textit{i.e.}, curvature) kernel
\begin{equation}
    L^2_{xy}=\left[\begin{array}{c@{\;\;}rc}
         1 & 1 & 1 \\
         1 & -8 & 1 \\
         1 & 1 & 1 
    \end{array}\right]\,,\label{eq:the_kernel}
\end{equation}
\noindent to identify flux peaks for localised fitting of Gaussian components \citep{hancock_2012}. Its flood-fill algorithm utilises two parameters, a seed threshold parameter, $\sigma_s$ (\textit{i.e.}, \texttt{seedclip}), above which to seed an island, and a flood threshold parameter, $\sigma_f$  (\textit{i.e.}, \texttt{floodclip}), above which to grow an island, such that, $\sigma_s\ge\sigma_f$ (see Table~\ref{tb:rms_isl_pars}). It then convolves the image with Equation~\ref{eq:the_kernel}, producing a curvature map, from which it implements Gaussian fits to local depressions (\textit{i.e.}, negative curvature bowls, corresponding to local flux-density maxima) within the islands. 

The implementation of the Aegean (version 2.2.4) module for Cerberus provides the box-car and flood-fill parameters, as per module design requirements.

Here we use the source component catalogue: \textit{i.e.}, source island information is not included in this version of Hydra.\footnote{\textit{i.e.}, the \texttt{-\,-island} flag is not set. For more details, see Aegean footnote $a$ in Table~\ref{tb:rms_isl_pars}.} Of particular interest are the integrated flux densities and fitted component sizes.

\subsection{Caesar}
\label{sc:caesar}
Caesar does its source finding using a flood-fill method to obtain blobs (\textit{i.e.}, ``islands''), from which child-blobs (or ``nested blobs'') are (optionally) extracted using elliptical-Gaussian based Laplacian (Equation~\ref{eq:the_kernel}) or $\chi^2$  filters \citep{riggi_2016,riggi_2019}. Compact sources, \textit{i.e.}, childless-blobs, are subtracted out to leave a residual map with extended sources that can be extracted either though a wavelet transform, saliency, hierarchical-clustering, or active-contour filter. 

The RMS and island parameters come in two sets, one for the parents, \texttt{seedThr} and \texttt{mergeThr}, respectively, and one for the children, \texttt{nestedBlobPeakZThr} and \texttt{nestedBlobPeakZMergeThr}, respectively. As we are optimising these parameters externally (through the PRD, Equation~\ref{eq:prd}), we set \texttt{compactSourceSearchNIters = 1} to prevent decrements in \texttt{seedThr} by \texttt{seedThrStep}. In our implementation, we also search for child-blobs (\textit{i.e.}, \texttt{searchNestedSources = true}), with their RMS and island parameters set to the same values as their parents \citep{riggi_2019}. For child-blob filtering we use the Laplacian method (\textit{i.e.}, \texttt{blobMaskMethod = 2}, with \texttt{fitSources = true}), and for source extraction we use the saliency filter method \citep[\textit{i.e.}, \texttt{extendedSearchMethod = 2}; for algorithms, see][]{riggi_2016}. 

It should be noted here that searching for child-blobs is not necessary for an image consisting of only point sources; however, for consistency, we prefer to have the same settings for both point and extended sources, for the purposes of comparing performance against other SFs. The only potential impact of this approach is in extra processing time.

Caesar also does background and RMS noise optimisation through any of the following metrics, $\mu/\sigma$, median/MADFM, biweight, and clipped median/$\sigma$. Consequently it does not require pre-tuning like PyBSDF, for example. Here we have chosen the median/MADFM metric (\textit{i.e.}, \texttt{bkgEstimator = 2}, with \texttt{useLocalBkg = true} and \texttt{useBeamInfoInBkg = true}), as it is similar to the pre-optimisation scheme option used by Typhon.

Table~\ref{tb:caesar_pars} summarises all of the internal settings of Cerberus's Caesar module. Note that we have also set some of the residual image processing flags, so as to remove all source types (\texttt{removedSourceType = -1}) with the appropriate thresholding (\textit{i.e.}, \texttt{residualZHighThr = seedThr} and \texttt{residualZThr = mergeThr}).

\begin{table}[htb!]
\caption{Caesar (Version 1.1.5) module settings. \label{tb:caesar_pars}}
\centering
\begin{tabular}{@{\;}llc@{\;}}
\hline\hline
Parameter & Value\\
\hline%
\texttt{useLocalBkg}                & \texttt{true}\\
\texttt{bkgEstimator}               & \texttt{2}\\
\texttt{useBeamInfoInBkg}           & \texttt{true}\\
\texttt{searchCompactSources}       & \texttt{true}\\
\texttt{compactSourceSearchNIterse} & \texttt{1}\\
\texttt{searchNestedSources}        & \texttt{true}\\
\texttt{extendedSearchMethod}       & \texttt{4}\\
\texttt{blobMaskMethod}             & \texttt{2}\\
\texttt{nestedBlobPeakZThr}         & \texttt{seedThr}\\
\texttt{nestedBlobPeakZMergeThr}    & \texttt{mergeThr}\\
\texttt{fitSources}                 & \texttt{true}\\
\texttt{computeResidualMap}         & \texttt{true}\\
\texttt{removeNestedSources}        & \texttt{true}\\
\texttt{removedSourceType}          & \texttt{-1}\\
\texttt{residualZHighThr}           & \texttt{seedThr}\\
\texttt{residualZThr}               & \texttt{mergeThr}\\
\texttt{saveResidualMap}            & \texttt{true}\\
\texttt{residualMapFITSFile}        & \texttt{residual.fits}\\
\hline\hline
\end{tabular}
\label{tb:caesar_module_settings}
\end{table}

The output source components are obtained from the source component catalogue, \textit{i.e.}, we do not include the source island catalogue in this version of Hydra. Of particular interest are the integrated flux densities and fitted component sizes.

\subsection{ProFound}
\begin{figure}[htb!]
\begin{center}
\includegraphics[width=\columnwidth]{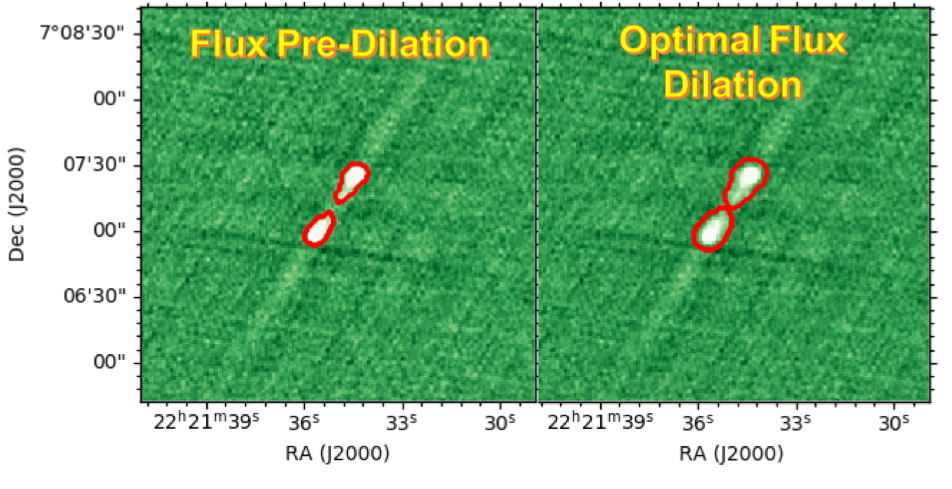}
\caption{Example of ProFound dilation process of a small $3^\prime\times3^\prime$ VLASS Epoch 1.1 Quick Look (QL) image cutout, centered at J222135+070712. The cutout was extracted from a QL image tile, available at NRAO (\url{https://science.nrao.edu}), and then processed using ProFound in R Studio. \label{fg:profound_dialation}}
\end{center}
\end{figure}

ProFound uses a watershed deblending process, wherein it systematically searches for the highest flux pixel and expands outwards and downwards in flux to some cutoff, creating a segment (\textit{i.e.}, ``island''), before proceeding to the next highest flux pixel, and so on \citep{robotham_2018}. The end result is the formation of ``flux-mountains'' (segments) with peaks and valleys (boundaries between segment groups). After the segments have been determined, it then dilates them until convergence is reached, as determined by a Kron/Petrosian-like dilation kernel (see \S~\ref{sc:source_finders}), 
while assigning overlapping segment fluxes to the ones with the most flux (\textit{e.g.}, Figure~\ref{fg:profound_dialation}). The segment formation threshold is determined by a \texttt{skycut} parameter, which corresponds to our RMS parameter, and the segment partitioning is determined by a \texttt{tolerance} parameter, which corresponds to our island parameter.

ProFound was designed for optical images and so there are some nuances when it comes to applying it to radio image data \citep[see][]{hale_2019}. Table~\ref{tb:profound_settings} summarises all of the settings internal to the ProFound module,\footnote{See also ProFound footnote $c$ in Table~\ref{tb:rms_isl_pars}.\label{fn:profound}} used herein. In addition, a considerable amount of wrapper code was required so as to extract the appropriate ``radio catalogue like'' information from its internal hierarchical data structure. Of particular interest are the total flux density and component size, which are pixel sums and flux-weighted fits,\footnote{For the major axis, it is the \textit{``weighted standard deviation along the major axes (i.e., the semi-major first moment, so $\sim$2 times this would be a typical major axis Kron radius) in units of pix.''} \textsuperscript{\ref{fn:profound}}} respectively, for a given segment \citep{robotham_2018}. Consequently these are not directly comparable to FWHM measurements from Gaussian fits in other SFs.

\begin{table}[htb!]
\caption{ProFound (version 1.13.1, with R version 4.0.3) module default settings, where \texttt{box = c(100, 100)}. It should be noted that results can differ radically between versions of ProFound and R. Furthermore, the default settings mentioned in the documentation can differ considerably from what is actually in the source code. This table includes all settings that are deemed important for reproducing our results. \label{tb:profound_settings}}
\centering
{\footnotesize\begin{tabular}{@{\;}ll|ll@{\;}}
\hline\hline
Parameter & Value & Parameter & Value\\
\hline
\texttt{pixcut}        &  \texttt{3}           &  \texttt{iterative}     & \texttt{FALSE}     \\
\texttt{ext}           &  \texttt{2}           &  \texttt{doclip}        & \texttt{TRUE}      \\
\texttt{reltol}        &  \texttt{0}           &  \texttt{shiftloc}      & \texttt{FALSE}     \\
\texttt{cliptol}       &  \texttt{Inf}         &  \texttt{paddim}        & \texttt{TRUE}      \\
\texttt{sigma}         &  \texttt{1}           &  \texttt{verbose}       & \texttt{TRUE}      \\
\texttt{smooth}        &  \texttt{TRUE}        &  \texttt{plot}          & \texttt{FALSE}     \\
\texttt{SBN100}        &  \texttt{100}         &  \texttt{stats}         & \texttt{TRUE}      \\
\texttt{size}          &  \texttt{5}           &  \texttt{rotstats}      & \texttt{TRUE}      \\
\texttt{shape}         &  \texttt{"disc"}      &  \texttt{boundstats}    & \texttt{TRUE}      \\
\texttt{iters}         &  \texttt{6}           &  \texttt{nearstats}     & \texttt{TRUE}      \\
\texttt{threshold}     &  \texttt{1.05}        &  \texttt{groupstats}    & \texttt{TRUE}      \\
\texttt{magzero}       &  \texttt{0}           &  \texttt{group}         & \texttt{NULL}      \\
\texttt{pixscale}      &  \texttt{1}           &  \texttt{groupby}       & \texttt{"segim"}   \\
\texttt{redosegim}     &  \texttt{FALSE}       &  \texttt{offset}        & \texttt{1}         \\
\texttt{redosky}       &  \texttt{TRUE}        &  \texttt{haralickstats} & \texttt{FALSE}     \\
\texttt{redoskysize}   &  \texttt{21}          &  \texttt{sortcol}       & \texttt{"segID"}   \\
\texttt{box}           &  \texttt{box}         &  \texttt{decreasing}    & \texttt{FALSE}     \\
\texttt{grid}          &  \texttt{box}         &  \texttt{lowmemory}     & \texttt{FALSE}     \\
\texttt{type}          &  \texttt{"bicubic"}   &  \texttt{keepim}        & \texttt{TRUE}      \\
\texttt{skytype}       &  \texttt{"median"}    &  \texttt{watershed}     & \texttt{"ProFound"}\\
\texttt{skyRMStype}    &  \texttt{"quanlo"}    &  \texttt{pixelcov}      & \texttt{FALSE}     \\
\texttt{roughpedestal} &  \texttt{FALSE}       &  \texttt{deblendtype}   & \texttt{"fit"}     \\
\texttt{sigmasel}      &  \texttt{1}           &  \texttt{psf}           & \texttt{NULL}      \\
\texttt{skypixmin}     &  \texttt{prod(box)/2} &  \texttt{fluxweight}    & \texttt{"sum"}     \\
\texttt{boxadd}        &  \texttt{box/2}       &  \texttt{convtype}      & \texttt{"brute"}   \\
\texttt{boxiters}      &  \texttt{0}           &  \texttt{convmode}      & \texttt{"extended"}\\
\texttt{iterskyloc}    &  \texttt{TRUE}        &  \texttt{fluxtype}      & \texttt{"Raw"}     \\
\texttt{deblend}       &  \texttt{FALSE}       &  \texttt{app\_diam}     & \texttt{1}         \\
\texttt{df}            &  \texttt{3}           &  \texttt{Ndeblendlim}   & \texttt{Inf}       \\
\texttt{radtrunc}      &  \texttt{2}           &                         &                    \\
\hline\hline
\end{tabular}}
\end{table}

\subsection{PyBDSF}
\label{sc:pybdsf}
PyBDSF identifies islands by collecting pixels greater than a given flux threshold, \texttt{thresh\_pix},\footnote{\textit{Re.} \texttt{process\_image(\ldots)} of footnote $d$ in Table~\ref{tb:rms_isl_pars}.} and then expands outwards from those pixels in octets above a given island-boundary threshold, \texttt{thesh\_isl}. Our RMS and island parameters correspond to former and latter thresholds, respectively. After the islands have been determined, it performs multiple Gaussian fits to each island \citep{hancock_2012}. 

We follow the same generic recipe as \cite{hale_2019} to accommodate extended sources, as outlined in Table~\ref{tb:pybdsf_pars}. The \texttt{atrous\_do} parameter selects the \`{a} trous wavelet decomposition \citep{holschneider_1989} module as one of several options for post processing. These include shapelet decomposition, \`{a} trous wavelet decomposition, 
PSF variation, polarisation, and spectral index modules. Setting \texttt{flag\_maxsize\_bm = 100} along with \texttt{atrous\_do = True} allows for Gaussians greater than the beam size and of varying scales, respectively. Setting \texttt{mean\_map = "zero"} sets the background mean to zero, enhancing the detection of extended emission.

\begin{table}[htb!]
\caption{PyBDSF (version 1.9.1) module settings. \label{tb:pybdsf_pars}}
\centering
\begin{tabular}{@{\;}llc@{\;}}
\hline\hline
Parameter & Value\\
\hline%
\texttt{atrous\_do}        & \texttt{True}\\
\texttt{flagging\_opts}    & \texttt{True}\\ 
\texttt{flag\_maxsize\_bm} & \texttt{100}\\ 
\texttt{mean\_map}         & \texttt{"zero"}\\ 
\texttt{interactive}       & \texttt{False}\\ 
\texttt{quiet}             & \texttt{False}\\ 
\hline\hline
\end{tabular}
\end{table}

The PyBDSF module also provides the \texttt{rms\_box} tuple, so that the RMS box and step sizes can be optimised by Typhon. 

The output source components are obtained from the source catalogue: \textit{i.e.}, \texttt{catalog\_type = 'srl'} \textit{via} the \texttt{write\_image(\ldots)} command. The catalogue does not include empty islands (\textit{i.e.}, \texttt{incl\_empty = False}). Of particular interest are the total flux densities and component sizes, which are expressed as integrated Stokes I and FWHM's, respectively.

\subsection{Selavy}
\label{sc:selavy}
Selavy is a ``single-pass'' raster-scan, or thresholding,  type SF \citep[see][]{lutz_1980}, with Duchamp \citep{whiting_2012b}, a 3D SF, at its heart \citep{whiting_2012}. Here we are interested in its 2D spatial search features (\textit{i.e.}, \texttt{searchType = spatial}). The algorithm works downwards by growing regions of detection through a threshold parameter \citep[see Figure~3 of][]{whiting_2012}, \texttt{snrCut}, our RMS parameter, after which they can be further extended downwards and outwards through an optional \texttt{growthCut} parameter, our island parameter. Further post processing options are available, such as producing components from multi-Gaussian fitting: \textit{i.e.}, \texttt{doFit = True}, to turn the fitting option on, \texttt{fitTypes = [full]}, to fit all degrees of freedom, and \texttt{numGaussFromGuess = True}, to provide an initial guess from the number of distinct peaks found within a given region during thresholding.\footnote{See Selavy footnote $e$ in Table~\ref{tb:rms_isl_pars}.}

Selavy also has various options for background estimates, such as typical $\mu$/$\sigma$ or more robust median/MADFM based statistics \citep{whiting_2012b}. We use a variable sliding box method (\texttt{VariableThreshold = True}) with robust statistics (\texttt{flagRobustStats = True}) and \texttt{Selavy.VariableThreshold.boxSize} = (\texttt{rms\_box}-1)/2, where \texttt{rms\_box} is determined by Typhon.

Table~\ref{tb:selavy_pars} summarises all of the internal settings of Cerberus's Selavy module.

\begin{table}[htb!]
\caption{Selavy (version 1.1.0) module settings. \label{tb:selavy_pars}}
\centering
\begin{tabular}{@{\;}llc@{\;}}
\hline\hline
Parameter & Value\\
\hline%
\texttt{Selavy.imagetype}                 & \texttt{fits}\\
\texttt{Selavy.flagLog}                   & \texttt{True}\\
\texttt{Selavy.flagDS9}                   & \texttt{True}\\
\texttt{Selavy.Fitter.doFit}              & \texttt{True}\\
\texttt{Selavy.Fitter.fitTypes}           & \texttt{[full]}\\
\texttt{Selavy.Fitter.numGaussFromGuess}  & \texttt{True}\\
\texttt{Selavy.searchType}                & \texttt{spatial}\\
\texttt{Selavy.VariableThreshold}         & \texttt{True}\\
\texttt{Selavy.flagRobustStats}           & \texttt{True}\\
\texttt{Selavy.flagGrowth}                & \texttt{True}\\
\hline\hline
\end{tabular}
\end{table}

The output source components are obtained from the source component catalogue: \textit{i.e.}, we do not include the source island catalogue in this version of Hydra. Of particular interest are the integrated flux densities and fitted component sizes.

\end{appendix}

\bibliographystyle{pasa-mnras}
\bibliography{references}

\end{document}